\documentclass[10pt, journal]{IEEEtran}

\usepackage{algorithm} 
\usepackage{algorithmic}
\usepackage{booktabs}
\usepackage{clrscode3e}
\usepackage{hyperref}
\usepackage{multirow}
\usepackage{graphicx}
\usepackage{subfigure}
\usepackage{textcomp}
\usepackage{xcolor}
\usepackage{acronym} 
\usepackage{amsfonts}
\usepackage{amssymb}
\usepackage{amsthm,amsmath}
\usepackage{mathrsfs}
\usepackage{indentfirst}
\usepackage{cases}
\usepackage{bm}
\usepackage{color}
\usepackage{cite}
\usepackage{stfloats}
\usepackage{url}
\usepackage{verbatim}
\usepackage{array}
\usepackage{amsmath,amsfonts}
\usepackage{threeparttable}
\acrodef{RL}{Reinforcement Learning}
\acrodef{SCSR}{Shared-account Cross-domain Sequential Recommendation}
\acrodef{RL-ISN}{\textbf{R}einforcement \textbf{L}earning-enhanced \textbf{I}nformation \textbf{S}haring \textbf{N}etwork}
\acrodef{ISN}{Information Sharing Network}
\acrodef{CR}{Cross-domain Recommendation}
\acrodef{CSR}{Cross-domain Sequential Recommendation}
\acrodef{GR}{Group Recommendation}
\acrodef{MDP}{Markov Decision Process}
\acrodef{RL-DF}{Reinforcement learning-enhanced Domain Filter}
\acrodef{GRU}{Gated Recurrent Unit}
\acrodef{MLP}{Multi-Layer Perceptron}
\acrodef{SAM}{Shared-Account Modeling}
\acrodef{DA-GCN}{Domain-Aware Graph Convolutional Network}
\acrodef{BCR}{Basic Cross-domain Recommender}
\acrodef{UIN}{User Identification Network}
\acrodef{GNN}{Graph Neural Network}
\acrodef{GCN}{Graph Convolutional Network}
\acrodef{GCNs}{Graph Convolutional Networks}
\acrodef{SCRM}{Self-attention-based Cross-domain Recommendation Machine}
\acrodef{SR}{Sequential Recommendation}
\acrodef{RNNs}{Recurrent Neural Networks}
\acrodef{RNN}{Recurrent Neural Network}
\acrodef{CNN}{Convolutional Neural Network}
\acrodef{MRR}{Mean Reciprocal Rank}
\acrodef{CF}{Collaborative Filtering}
\acrodef{SA}{Self-Attention}
\acrodef{CR}{Cross-domain Recommendation}
\acrodef{DA-GCN}{Domain-Aware Graph Convolutional Network}
\acrodef{TiDA-GCN}{Time Interval-enhanced Domain-Aware Graph Convolutional Network}
\acrodef{CDS}{Cross-Domain Sequence}
\acrodef{CSR}{Cross-domain Sequential Recommendation}
\acrodef{SAR}{Shared-Account Recommendation}

\newcommand{\orcid}[1]{\href{https://orcid.org/#1}{\includegraphics[width=10pt]{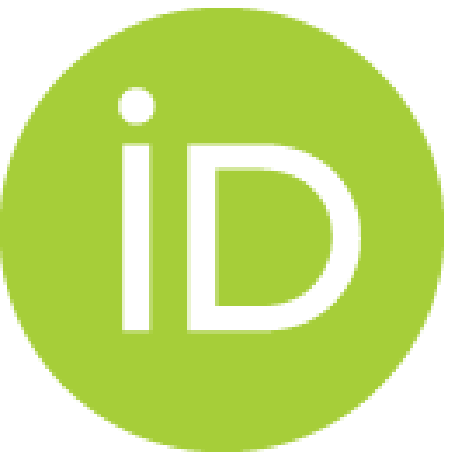}}}

\begin{document}
\title{Time Interval-enhanced Graph Neural Network for Shared-account Cross-domain Sequential Recommendation}
%
%
%
%
\author{Lei~Guo\orcid{0000-0002-9408-7594},
        Jinyu~Zhang\orcid{0000-0001-8842-5828},
        Li~Tang\orcid{0000-0003-0741-9157},
        Tong~Chen\orcid{0000-0001-7269-146X},
        Lei~Zhu\orcid{0000-0002-2993-7142} and
        Hongzhi~Yin\orcid{0000-0003-1395-261X}, ~\IEEEmembership{Senior Member,~IEEE}
\thanks{L. Guo, J. Zhang, L. Tang, and L. Zhu are with the School of information science and Engineering, Shandong Normal University, Jinan 250014, China.\protect\\
E-mail: leiguo.cs@gmail.com, jinyuz1996@outlook.com, litang96@126.com, leizhu0608@gmail.com}
\thanks{T. Chen and H. Yin are with the School of Information Technology \& Electric Engineering, The University of Queensland, St Lucia, QLD 4072, Australia.\protect\\
E-mail: tong.chen@uq.edu.au, h.yin1@uq.edu.au}

\thanks{Manuscript received XX XX, 2022; revised XX XX, 2022.\\
(Corresponding author: Hongzhi Yin.)}}

%
%


\markboth{IEEE TRANSACTIONS ON NEURAL NETWORKS AND LEARNING SYSTEMS,~VOL.~XX, NO.~X, X~2022}%
{GUO \MakeLowercase{\textit{et al.}}: TIME INTERVAL-ENHANCED GRAPH NEURAL NETWORK FOR SCSR}


\maketitle
%




\begin{abstract}
\ac{SCSR} task aims to recommend the next item via leveraging the mixed user behaviors in multiple domains. It is gaining immense research attention as more and more users tend to sign up on different platforms and share accounts with others to access domain-specific services. Existing works on \ac{SCSR} mainly rely on mining sequential patterns via \ac{RNN}-based models, which suffer from the following limitations: 1) \ac{RNN}-based methods overwhelmingly target discovering sequential dependencies in single-user behaviors. They are not expressive enough to capture the relationships among multiple entities in \ac{SCSR}. 2) All existing methods bridge two domains via knowledge transfer in the latent space, and ignore the explicit cross-domain graph structure. 3) None existing studies consider the time interval information among items, which is essential in the sequential recommendation for characterizing different items and learning discriminative representations for them. In this work, we propose a new graph-based solution, namely TiDA-GCN, to address the above challenges. Specifically, we first link users and items in each domain as a graph. Then, we devise a domain-aware graph convolution network to learn user-specific node representations. To fully account for users’ domain-specific preferences on items, two effective attention mechanisms are further developed to selectively guide the message passing process. Moreover, to further enhance item- and account-level representation learning, we incorporate the time interval into the message passing, and design an account-aware self-attention module for learning items’ interactive characteristics. Experiments demonstrate the superiority of our proposed method from various aspects.
\end{abstract}


\section{Introduction \label{sec:introduction}}

\IEEEPARstart{C}{ross-domain} Sequential Recommendation (CSR) task aims at recommending the next item by leveraging a user’s historical interactions from multiple domains. It is gaining immense research attention as users tend to sign up on different platforms to access domain-specific services, such as video watching and news pushing platforms. Ma et al.~\cite{ma_mixed_TKDD} propose a mixed information flow network to simultaneously consider the flow of behavioral information and knowledge across domains. Li et al.~\cite{Li_dual_KDD_2021} devise a dual learning mechanism to transfer information between two related domains in an iterative manner for CSR. However, existing studies mainly focus on making recommendations for single users, and they fail to consider the common fact that people share accounts with others in many real applications. 
The shared account is a kind of account that is shared by a group of users with similar interests or who have intense social relationships.
For example, friends and family members tend to share an account for watching movies (e.g., Netﬂix) and online shopping (e.g., Amazon).

In this work, we study CSR in an emerging yet challenging scenario, i.e., \ac{SCSR}, where multiple individual users share a single account and their interaction behaviors are recorded in multiple domains~\cite{ma2019pi}.
The shared account is different from the user overlapping concept in the cross-domain recommendation, since user overlapping means there are common users between two domains (e.g., the source and target domains), and based on which, we can align two domains.
Previous studies on making recommendations to shared-accounts~\cite{bajaj2016experience} are proposed to capture the user relationships under the same account with latent representations. None of them consider the cross-domain context, and thus they are inapplicable to \ac{SCSR}. Compared to CSR, \ac{SCSR} is a more challenging task, because: 1) Users’ diverse interests are mixed in an interaction sequence~\cite{ma2019pi,guo_DAGCN_2021}. Under such circumstances, directly treating an account as a virtual user may get sub-optimal results, since this strategy will easily result in the loss of the user-specific information. 2) The existence of shared accounts will amplify the noise in the interaction data and impede uncovering users’ preferences from the data in two domains.

Recently, several methods have been developed for \ac{SCSR}. Among them, $\pi$-Net~\cite{ma2019pi} is the first work for \ac{SCSR}. It solves a parallel sequential recommendation problem with a \ac{GRU}-based information-sharing network. Another related work is the PSJNet method~\cite{ren2019net}. It improves $\pi$-Net by using a split-join strategy on \ac{RNN}. However, these existing works still have the following limitations: 
1) Existing methods for \ac{SCSR} are all \ac{RNN}-based methods, which overwhelmingly target discovering sequential dependencies, and have limited capability for capturing the complex relationships among associated entities (i.e., users and items) in both domains. As a result, this limits the expressiveness of learned user and item representations. 
2) All existing methods try to transfer the domain knowledge in the latent space, and overlook the explicit structural information (e.g., item-user-item paths) linking two domains. Though traditional cross-domain recommenders~\cite{zhuang2018cross} that focus on knowledge transfer can partially solve this problem, most of them implicitly share the cross-domain knowledge in the latent space, and the explicit structural information bridging the two domains is not sufficiently explored. Zhao et al.~\cite{chengzhao2019ppgn} propose a graph-based cross-domain recommendation method. But their work ignores the important sequential information and relies on explicit user ratings that are not usually available in both domains. Moreover, their method cannot handle the entangled user preferences on shared accounts.
3) None of the existing studies on \ac{SCSR} considers the time interval information among items. We argue that only regarding interaction histories as ordered sequences, without the time intervals, will result in inaccurate sequence modeling and sup-optimal recommendations in sequential recommendation~\cite{bai2019ctrec,wang2020make,TamWTYH17,zhang2021neural}.
Intuitively, different types of items have different frequencies and periodicities when being consumed, and therefore, they should have different influences on the next item.
For instance, users tend to consume fast-moving consumer goods in a short period, and buy electronic products at long intervals. Such additional knowledge provides strong predictive signals when inferring the next item to be visited.

To address the above limitations in \ac{SCSR}, in our previous work, we devise a graph-based solution, namely \ac{DA-GCN}, and publish it as a conference paper in IJCAI 2021~\cite{guo_DAGCN_2021}. Specifically, to incorporate the complicated interaction relationships (limitation 1), \ac{DA-GCN} constructs a \ac{CDS} graph in advance to link users and items in different domains. To model the structural information across domains (limitation 2), \ac{DA-GCN} devises a domain-aware message-passing strategy in a \ac{GCN} to learn user-specific account and item representations from their connected neighbors.
However, \ac{DA-GCN} solely focuses on solving the limitations 1 and 2, it does not address the limitation 3. Moreover,
as DA-GCN only leverages one representative item as the sequence-level representation, the users' preferences over other items and their interactive characteristics are both ignored. Though the average pooling strategy can take all the items into consideration, it treats items equally and cannot learn their different contributions to the sequence.

To further deal with the limitation 3, and capture the interactive characteristics that are ignored by \ac{DA-GCN}, we extend \ac{DA-GCN} by devising a time interval enhanced graph neural network, namely \ac{TiDA-GCN}. 
To be more concrete, to model the time interval information between items (limitation 3), we further develop a time interval-aware message passing strategy, where the sequential orders and the relative time intervals between any two items are both exploited. Afterward, to learn items’ interactive characteristics from the sequence level, we introduce an account-aware self-attention network to aggregate all the items attentively. Consequently, we can model the multifaceted interactions and transfer the fine-grained domain knowledge by considering the time interval-aware structure information.
Compared with the conference version, we have re-written the Related Work
(Section~\ref{related_work}), the Methodologies (Section~\ref{sec:methodologies}), and the Experiments
(Section~\ref{Experimental_setup}, Section~\ref{results}, and Section~\ref{analysis}) accordingly.

The main contributions of this work are summarized as follows:
\begin{itemize}
\item We investigate an emerging yet challenging recommendation task, namely \ac{SCSR}. After pointing out the defects of existing solutions, we bring a new take on the \ac{SCSR} problem with a time interval enhanced graph neural network.
\item We develop a time interval message passing strategy by further considering the time interval information on the basis of \ac{DA-GCN} to enhance the item representation learning.
\item We devise an account-aware self-attention mechanism to model items' interactive characteristics in the account representation learning.
\item We conduct extensive experiments on two real-world datasets, and demonstrate the superiority of our proposed method compared with several state-of-the-art baselines.
\end{itemize}
\section{Related Work\label{related_work}}
This section surveys four types of related works: Cross-domain Recommendation, Shared-account Recommendation, \ac{GCN}-based Recommendation, and Time Interval-aware Sequential Recommendation.

\subsection{Cross-domain Recommendation}

\ac{CR} that concerns data from multiple domains has proven useful for dealing with cold-start~\cite{abel2013cross,lu2020meta, Zhang2019crossdomain} and data sparsity issues~\cite{WangNL20, gao2019cross}. In this work, we limit the number of domains to two, which is a common practice in most \ac{CR} methods~\cite{LiuSCZ21,ZhuWCLZ20, Zhang2019crossdomain,hu2018conet}. Existing studies with this limitation can be categorized into shallow methods and deep learning-based methods.

\textbf{Shallow Methods.}
In this type, two main ways are developed for addressing \ac{CR}. One is to aggregate knowledge between two domains. For example, Loni et al.~\cite{loni2014cross} first define a domain as a type of item, and then apply factorization machines to \ac{CR} by incorporating user interaction patterns that are related to particular types of items. Chen et al.~\cite{chen2017tlrec} tend to connect different domains by embedding all users and items into the same low dimensional potential space and learning parameters via normalizing the proximity between them. Man et al.~\cite{man2017cross} study \ac{CR} from an embedding and mapping perspective, and capture the underlying mapping relationship between the user/item feature vectors in different domains by devising a \ac{MLP}-based mapping function.
The other focuses on transferring knowledge from the source domain to the target domain. For example, Hu et al.~\cite{hu2013personalized} leverage a tensor-based factorization method to share latent features between different domains. 
Doan et al.~\cite{doan_transcorssCF} investigate a transition-based cross-domain collaborative filtering method to model both within and between domain transitions of user historical sequential behaviors.

\textbf{Deep Learning-based Methods.}
Deep learning-based methods are well suited to transfer learning, as they can learn high-level abstractions among different domains~\cite{hu2018conet, chengzhao2019ppgn, liu2020transfer}. For example,
Hu et al.~\cite{hu2018conet} investigate the benefit brought by the deep transfer learning method, and propose a collaborative cross-network to enable dual knowledge transfer across domains via introducing cross-connections from one base network to another and vice versa.
Zhao et al.~\cite{chengzhao2019ppgn} construct a cross-domain preference matrix to model the interactions of different domains as a whole and design a preference propagation network on the basis of \ac{GCN}.
Zhang et al.~\cite{Zhang2019crossdomain} investigate the dual-target \ac{CR} task with partially overlapping entities, and present a kernel-induced knowledge transfer-based recommendation method to address this issue.
Wang et al.~\cite{WangNL20} focus on the data sparsity and data imbalance issues in cross-domain recommender systems, and propose a discriminative adversarial network to transfer the latent representations from a source domain to a target domain in an adversarial way.
However, the above works are mainly focused on users' static interactions, and user behaviors' sequential patterns and the shared-account characteristic are not considered.
$\pi$-Net~\cite{ma2019pi} and PSJNet~\cite{ren2019net} are two recently proposed methods for \ac{SCSR}, but their studies are all based on RNNs, which are neither expressive enough to capture the multiple associations nor can model the structure information that bridges two domains.

There are also a few studies focusing on the multi-domain \ac{CR} scenario~\cite{ZhaoLXDS20, CuiWZZ20, ZhangLHMCLT20}, whose main purpose is to improve recommendations for all domains simultaneously via capturing the informative domain-specific features from all of them. Compared with the scenarios of two domains, there is a new challenge in the multi-domain \ac{CR}~\cite{ZhuW00L021}, that is, the recommendation performance in some domains may decline as more domains join in (especially sparser domains).
This is also termed the negative transfer problem, i.e., the transferred knowledge may negatively affect the recommendation accuracy in the target domain. Though the recommendations in two domains may also face negative transfer issues, this problem in the multi-domain \ac{CR} scenario is amplified, since the information/knowledge
in each domain should be transferred to other domains more than once. We will extend our problem setting to multiple domains in our future work.

Another kind of \ac{CR} is the single-target multi-domain recommendation task~\cite{BiSYWWX20,FuPWXL19}, which tends to transfer the knowledge of common users among multiple domains to make recommendations for a different domain. But as they are designed to make single-target recommendations for one domain, they differ significantly from our \ac{SCSR} task.

\subsection{Shared-account Recommendation}
The task of recommending items for a shared account is to make recommendations by modeling the mixture of user behaviors, which usually performs user identification first, and then makes recommendations~\cite{zhao2016passenger,jiang2018identifying}. For example, Wang et al.~\cite{wang2014user} are the first to study user identification from a novel perspective, in which they employ user preference and consumption time to indicate a user's identity.
Bajaj et al.~\cite{bajaj2016experience}
focus on decomposing online TV accounts into distinct personas by a similarity-based channel clustering method, and then make recommendations to individualize the experience for each persona.
Jiang et al.~\cite{jiang2018identifying} introduce an unsupervised framework to judge users within the same account and then learn the preferences for each of them. In their method, a heterogeneous graph is exploited to learn the relationship between items and the metadata. However, all the above methods are focused on identifying latent users in a single domain, rather than a cross-domain scenario.
Wen et al.~\cite{wen2021miss} address the shared-account challenge in a session-aware recommendation task, where a multi-user identification module drawing on the attention mechanism to distinguish behaviors of different users is proposed.
The recently proposed methods $\pi$-Net~\cite{ma2019pi} and PSJNet~\cite{ren2019net} argue that this task can be treated in an end-to-end manner, and can also be improved by simultaneously considering the cross-domain information. In their approaches, each account is assumed to have $H$ latent users, and the account representation is learned by merging all the user-specific representations of latent users within it.

\subsection{GCN-based Recommendation}
The development of graph neural networks~\cite{wang2019ngcf,wang_origin_2019,Yu2022graph} has attracted a lot of attention to exploring graph-based solutions for recommender systems~\cite{guo2021hierarchical,wang2021graph,Cai2018Graph}. For example, Ying et al.~\cite{ying2018grasage} develop a data-efficient graph convolutional network by combining random walks and graph convolutions to generate embeddings for items. Wang et al.~\cite{wang2019ngcf} exploit high-order proximity by propagating embeddings on the user-item interaction graph to simultaneously update the representations for all users and items efficiently by implementing the matrix-form rule. He et al.~\cite{He2020lightGCN} find two common designs in \ac{GCN}, i.e., feature transformation and nonlinear activation, contribute little to the collaborative filtering method, and propose a simplified version of GCN to make it more concise and appropriate for the recommendation. 
To explore \ac{GCN} for sequential recommendations, Wu et al.~\cite{shuwu2019gnn} model session sequences as graph-structured data to capture the complex transitions of items, which are difficult to be revealed by previous conventional sequential methods.
Qiu et al.~\cite{qiu_exploiting_2020} address the session-based recommendation problem via proposing a novel full graph neural network to model the item dependencies between different sessions. Chang et al.~\cite{chang2021sequential} 
deal with the challenges that users' behaviors are often noisy signals and their preferences usually change over time by proposing a graph neural network. To be more specific, they first integrate different types of preferences into clusters, and then perform cluster-aware and query-aware graph convolutional propagation on the constructed graph.
However, none of the GCN-based methods can be directly applied to \ac{SCSR} as they either fail to capture the critical sequential information~\cite{wang2020intention} or focus on the recommendation problem in a single domain.

\subsection{Time Interval-aware Sequential Recommendation}
Time interval as one of the important factors reflecting users' dynamic preferences has been widely studied in the sequential recommendation scenario. Based on different research perspectives, existing studies can be categorized into user- and item-oriented modeling methods.
User-oriented methods~\cite{ChenGYYXYY21, WangJWPW18} are mainly focused on exploring the time intervals from the user side to study users' continuous preferences.
For example, Chen et al.~\cite{ChenGYYXYY21} propose a self-modulating attention mechanism to model the complex evolution of user preferences over time, which enables eliminating the impact of the items that the user might not be interested in during certain time intervals.
Wang et al.~\cite{WangJWPW18} combine the temporal factor with a personalized tag recommendation method by adjusting the users' preference weights according to the time intervals between their past behaviors and the current time. 
Though the temporal dynamics on the user side are well investigated, the time gaps between items are often ignored and assumed they are of equal contributions.

To tackle this limitation, recent works have also focused on investigating the importance of time spans between items~\cite{ChenZYWW21, JiWWCC20, LiWM20TIASSR, JinC0HF022, ZhangBWW20, ZhuLLWGLC17, FanLZX0Y21, ChenZYWW21time}.
For example, Zhu et al.~\cite{ZhuLLWGLC17} model time intervals between users' actions by equipping LSTM with specifically designed time gates to better capture both users' short- and long-term interests, so as to improve the recommendation performance.
TGSRec~\cite{FanLZX0Y21} is a recent time interval-aware sequential recommendation method that explicitly models collaborative signals and time gaps among items via a collaborative attention network, where the time span is represented as a kernel value of the corresponding encoded time embeddings.
Chen et al.~\cite{ChenZYWW21time} first segment the whole timeline into a specified number of equal-length time slices, and then construct a user-item graph for each time slice to learn user and item representation.
Li et al.~\cite{LiWM20TIASSR} argue that user interaction sequences should be modeled as a sequence with different time intervals, and devise a time-aware self-attention mechanism to capture the time intervals. But both of them only explore the time intervals in sequential recommendations within a single domain, and none of them investigate it in \ac{SCSR}.

\subsection{Differences}
The proposed \ac{TiDA-GCN} method in this paper significantly extends and advances the methodology in the conference version (\ac{DA-GCN}).
In particular, our previous work~\cite{guo_DAGCN_2021} only leverages the structural information to learn account and item representations with a domain-aware graph convolutional network, while this work further incorporates the time interval information by devising an interval-aware message passing strategy to consider items' different attractions to users.
Furthermore, this work develops an account-aware self-attention network to model items' interactive characteristics, ignored by \ac{DA-GCN}, at the sequence level to enhance the account representation learning.

Compared with the time interval modeling methods, our method also has significant differences from them. For example, the \ac{TiDA-GCN} method is different from~\cite{ChenZYWW21time}, since we focus on designing time interval-aware message passing strategies for representation learning in \ac{GCN}s, rather than directly learning items' dynamic representations through RNNs.
Another related work to ours is~\cite{LiWM20TIASSR}, which models time intervals in an interaction sequence as the relation between items. But this work mainly explores the time intervals for the self-attention mechanism, and proposes an extension to self-attention, while ours mainly focus on improving \ac{GCN}s via devising time interval-aware message passing strategies.
Moreover, all the existed studies only explore the time intervals in sequential recommendations within a single domain, and none of them investigate it in \ac{SCSR}, which is the main task of our work.
\section{Methodologies \label{sec:methodologies}}
In this section, we first provide the preliminaries of this work and formulate the \ac{SCSR} task that we intend to address. Then, we introduce how we leverage the \ac{GCN} technique to meet the challenges, and propose a \acf{TiDA-GCN} as our solution. For clarity, we also introduce our previous work \ac{DA-GCN}~\cite{guo_DAGCN_2021} that learns account and item representations without the time interval and sequence-level information as a comparison.

\begin{figure*}
    \centering
    \includegraphics[width=15cm]{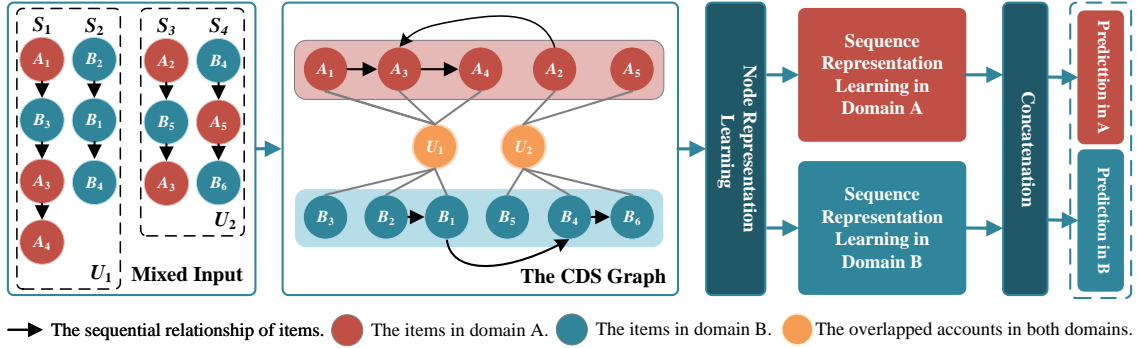}
    \caption{The workflow of \ac{TiDA-GCN}, where $S_1$ to $S_4$ denote the historical behavior sequences of two shared-accounts ($U_1$ and $U_2$) in both domains.}
    \label{fig:DA_overview}
\end{figure*}

\subsection{Preliminaries}
Suppose $\mathcal{U} = \left\{U_1,U_2,...,U_k,...\right\}$ is the set of shared-accounts in both domains, where $U_k \in\mathcal{U}$ $(1\leq k\leq n)$ denotes a single account index in $\mathcal{U}$ ($n$ is the size of $\mathcal{U}$).
Let us define $ S_A = \{ A_1, A_2, \dots, A_i, ...\}$ and $ S_B = \{ B_1, B_2, \dots,  B_j, ...\} $ as the sequences from a shared account in domain A and B respectively, where $A_i \in\mathcal{A}$ $(1\leq i\leq p)$ represents a single item index in domain A, and $\mathcal{A}$ denotes the whole item set in domain A ($p$ is the number of items in $\mathcal{A}$). Analogously, $B_j \in  \mathcal{B}$ $(1\leq j\leq q)$ denotes a single item index in domain B, and $\mathcal{B}$ is the whole item set in B ($q$ is the number of items in $\mathcal{B}$). 
Given $S_A$ and $S_B$, our \ac{SCSR} task tends to recommend the next item for an account by modeling the behavior sequences from it. We define the recommendation probabilities for all candidate items in domains A and B as:
\begin{align}
& P(A_{i+1}|S_A, S_B)\sim f_A(S_A, S_B), \\
& P(B_{j+1}|S_B, S_A)\sim f_B(S_B, S_A),
\end{align}
where $P(A_{i+1}|S_A, S_B)$ denotes the probability of recommending the item $A_{i+1}$ that will be consumed in domain A, given $S_A$ and $S_B$. $f_A(S_A,S_B)$ represents the learned function that is utilized to estimate $P(A_{i+1}|S_A,S_B)$. For domain B, we apply similar definitions to $P(B_{j+1}|S_B, S_A)$ and $f_B(S_B, S_A)$.

\subsection{Overview of TiDA-GCN}
The key idea of \ac{TiDA-GCN} is to explore the multiple associations among users and items, and the structure-aware domain knowledge via devising a domain-aware graph convolutional network, which not only models users' interests from the source domain A, but also transfers the information to the target domain B to improve the recommendation performance for B, and vice versa.
Fig.~\ref{fig:DA_overview} shows
the workflow of our recommendation framework, which consists of five components:
1) The \ac{CDS} graph is constructed from the mixed behavior sequences to model four types of sophisticated associations among accounts and items, that is, sequential transitions among items in domain A, sequential transitions among items in domain B, account-item interactions in domain A, and account-item interactions in domain B. Note that, for the interactions between sequences, we handle them by constructing a global interaction graph according to items' sequential transition relationships in each domain, that is, for any two items, as long as they are adjacent in the sequences, we will assume they have a sequential relationship and connect them in the \ac{CDS} graph. That is to say, items between sequences are connected because they are adjacent in a sequence. Otherwise, they will not be connected.
2) Account representation learning that leverages a domain-aware message passing strategy to learn node representations with latent users (suppose there are $H$ latent users sharing an account), where a domain-aware attention mechanism is developed to capture users' distinct preferences over different items on both source and target domains (the details can be seen in Section~\ref{account_representation}).
3) Item representation learning via exploiting a sequence-aware attention mechanism on the item side to access the varying relevance of linked users/items to the target item. Moreover, as item pairs with different time intervals may have different needs for users, such as fast-moving consumer goods and electronic products, the time intervals among them provide us important signals for item distinction. Hence, we further enhance the representation learning process by incorporating the time interval information into the message aggregation strategy (the details can be seen in Section~\ref{item_embedding}). 
4) Sequence representation learning that aims at investigating the sequential preferences of accounts at the sequence level, where a self-attention mechanism is leveraged (the details can be seen in Section~\ref{sequence_embedding}).
5) The model prediction layer conducts item recommendation by a joint training strategy to simultaneously optimize the cross-entropy objective function in both domains (the details can be seen in Section~\ref{prediction_layer}).
Note that, we can achieve our preliminary work \ac{DA-GCN} by disabling the time interval information and using the max pooling operation for sequence embedding.

\begin{figure*}
    \centering
    \includegraphics[width=14.5cm]{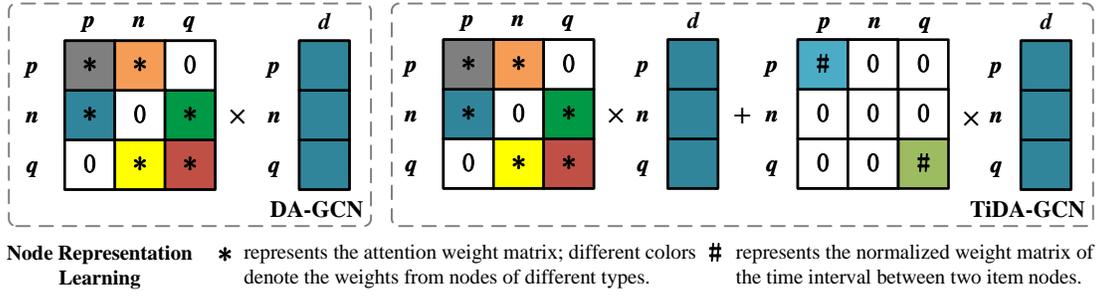}
    \caption{The node representation learning components in \ac{DA-GCN} and \ac{TiDA-GCN}.
    }
    \label{fig:node_representation_learning}
\end{figure*}

\subsection{Account Representation Learning with Latent Users\label{account_representation}}
In reality, as an account is usually shared by multiple users, and users under the same account often have different preferences, uniformly treating accounts as virtual users is inappropriate. Intuitively, the users sharing an account should have diversified representations, due to their multifaceted personalities and preferences. Motivated by this observation, we assume $H$ latent users ($U_{k,1}, U_{k,2}, ...., U_{k,h}, ..., U_{k,H}$) under each account $U_k$, and denote the embedding of the $h$-th latent user $U_{k,h}$ as $\bm{e}_{U_{k,h}} \in \mathbb{R}^d$, which is learned by aggregating the messages passed from items in both source and target domains (this process is shown in Fig.~\ref{fig:node_representation_learning}). 
Note that, as in real-world scenarios, we have no further information to distinguish the exact users sharing an account, such as family TV accounts; we subsume each account has $H$ latent users, and leverage $H$ as a hyper-parameter by fine-tuning it in different application scenarios.

\textbf{Message Passing.}
Let $A_i\in \mathcal{N}^{U_{k,h}}_{A_i}$ and $B_j\in \mathcal{N}^{U_{k,h}}_{B_j}$ be the connected items of $U_{k,h}$ in domain A and B respectively, where $\mathcal{N}^{U_{k,h}}_{A_i}$ denotes the neighbor set of $U_{k,h}$ in domain A, and $\mathcal{N}^{U_{k,h}}_{B_j}$ denotes the neighbor set of $U_{k,h}$ in domain B. 
Then, we define the representation ($\bm{m}_{U_{k,h} \leftarrow A_i}$) of the passed message from $A_i$ to $U_{k,h}$ in domain A as:
\begin{equation}
\bm{m}_{{U_{k,h}} \leftarrow {A_i}}=\gamma_{A_i}^{U_{k,h}} (\bm{W}_1\bm{e}_{A_i}+\bm{W}_2(\bm{e}_{A_i}\odot \bm{e}_{U_{k,h}})),
\end{equation}
where $\bm{W}_1, \bm{W}_2\in \mathbb{R}^{d^\prime \times d} $ are the trainable weight matrices, $\bm{e}_{A_i} \in \mathbb{R}^d$ and $\bm{e}_{U_{k,h}}$ are the embedding vectors of item $A_i$ and user $U_{k,h}$ respectively.
The element-wise product $\bm{e}_{A_i}\odot \bm{e}_{U_{k,h}}$ denotes the information interaction between $A_i$ and $U_{k,h}$.
$\gamma_{A_i}^{U_{k,h}}$ is a learnable parameter that indicates how much information we pass from $A_i$ to $U_{k,h}$. The learning process of it is introduced in the following subsection.

Similar to domain A, we define the representation ($\bm{m}_{{U_{k,h}} \leftarrow {B_j}}$) of the passed message from item $B_j$ to the target user $U_{k,h}$ in domain B as:
\begin{equation}
\bm{m}_{{U_{k,h}} \leftarrow {B_j}}=\gamma_{B_j}^{U_{k,h}} (\bm{W}_1\bm{e}_{B_j}+\bm{W}_2(\bm{e}_{B_j}\odot \bm{e}_{U_{k,h}})).
\end{equation}
Besides, we further add a self-connection to $U_{k,h}$ to keep the information carried by the target user. The passed information ($\bm{m}_{{U_{k,h}} \leftarrow {U_{k,h}}}$) is defined as:
\begin{equation}
\bm{m}_{{U_{k,h}} \leftarrow {U_{k,h}}}=\gamma_{U_{k,h}}^{U_{k,h}} (\bm{W}_1\bm{e}_{U_{k,h}}),
\end{equation}
where $\gamma_{B_j}^{U_{k,h}}$ and $\gamma_{U_{k,h}}^{U_{k,h}}$ are the attentive weights indicating the strength of each passed message, whose values are also computed by the following attention mechanism.

\textbf{Domain-aware Attention Mechanism.}
To fully capture the distinct preferences of each latent user over different items on both domains, we devise a domain-aware attention mechanism to measure the importance of item neighbors to the target user. The importance of $A_i$ to $U_{k,h}$ in domain A is defined as:
\begin{equation}
s_{A_i}^{U_{k,h}}=f(\bm{e}_{U_{k,h}},\bm{e}_{A_i}),
\end{equation}
where $f(\cdot)$ is a pairwise similarity metric. In this work, the cosine similarity function is applied.

Then, we achieve the attentive weight on each item $A_i$ via the following normalization operation:
\begin{align}
 \gamma_{A_i}^{U_{k,h}} = \text{exp}(s_{A_i}^{U_{k,h}})/ (\sum_{{A_i}^\prime \in \mathcal{N}_{A_i}^{U_{k,h}}}^{}\text{exp}(s_{{A_i}^\prime}^{U_{k,h}}) \notag \\ { + \sum_{B_j^\prime \in \mathcal{N}_{B_j}^{U_{k,h}}}^{}\text{exp}(s_{{B_j}^\prime}^{U_{k,h}})+ s_{U_{k,h}}^{U_{k,h}}}),
\end{align}
where $s_{B_j}^{U_{k,h}}$ and $s_{U_{k,h}}^{U_{k,h}}$ weight the importance of each item $B_j$ to $U_{k,h}$ in domain B, and the self-connection within $U_{k,h}$ respectively (similar to the definition of $s_{A_i}^{U_{k,h}}$).
By replacing the corresponding user/item embeddings, we can further obtain $\gamma_{B_j}^{U_{k,h}}$ and $\gamma_{U_{k,h}}^{U_{k,h}}$ in a similar way.

\textbf{Message Aggregation.}
Thereafter, we can obtain the embedding ($\bm{e}_{U_{k,h}}$) of user $U_{k,h}$ by aggregating all the passed messages in both domains:
\begin{align}
& \bm{e}_{U_{k,h}} = \text{LeakyReLU}(\bm{m}_{{U_{k,h}} \leftarrow {U_{k,h}}} \notag \\
& \qquad 
+ \sum_{A_i\in \mathcal{N}^{U_{k,h}}_{A_i}}{\bm{m}_{U_{k,h} \leftarrow A_i}}+ \sum_{B_j \in \mathcal{N}^{U_{k,h}}_{B_j}}{\bm{m}_{U_{k,h} \leftarrow B_j}}),
\end{align}
where LeakyReLU is a nonlinear activate function. $\bm{m}_{{U_{k,h}} \leftarrow {U_{k,h}}}$ is the information from herself. $\bm{m}_{U_{k,h} \leftarrow A_i}$ and $\bm{m}_{U_{k,h} \leftarrow B_j}$ are the passed messages from domain A and B, respectively.

To this end, we achieve the account representation by merging the embeddings of all the latent users:
\begin{equation}
\bm{e}_{U_k} = \frac{1}{H}\sum^{H}_{h=1}{e}_{U_{k,h}},
\end{equation}
where $\bm{e}_{U_k}$ is the representation of account $U_k$.

\begin{figure*}
    \centering
    \includegraphics[width=17.5cm]{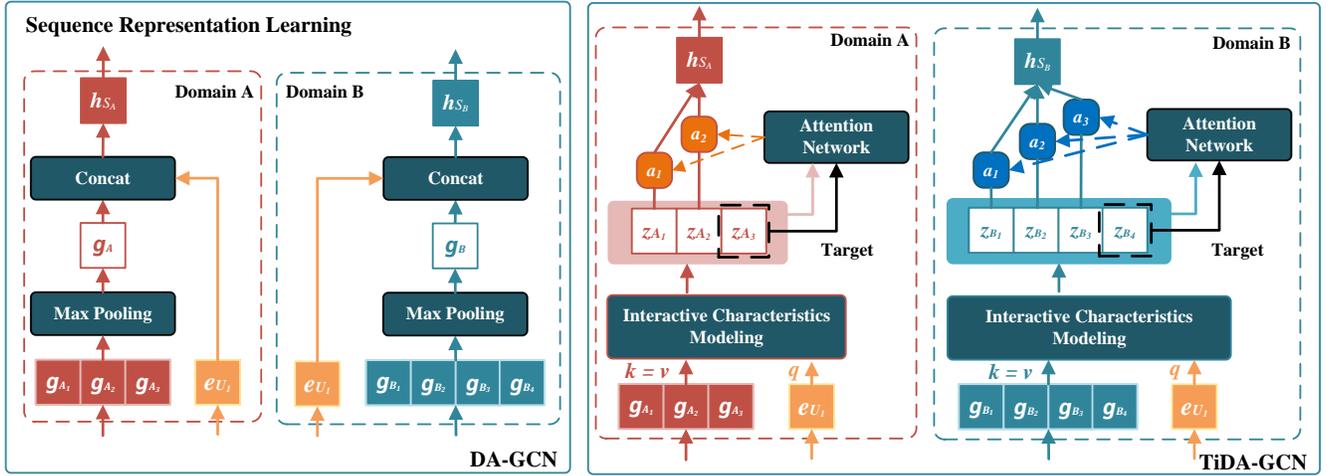}
    \caption{The sequence representation learning components in \ac{DA-GCN} and \ac{TiDA-GCN}.}
    \label{fig:SRL}
\end{figure*}

\subsection{Time Interval-aware Item Representation Learning\label{item_embedding}}
We model an item by aggregating the information from two types of nodes with connections, i.e., the connected users and items within the same domain. Moreover, as the item pairs with different time intervals usually reveal different attractions to users, such as users tend to buy fast-moving consumer goods in a short period, and buy electronic products at long intervals, modeling time intervals helps distinguish different items and capture fine-grained preferences.

\textbf{Message Passing.}
Let $U_{k,h}\in \mathcal{N}^{A_i}_{U_{k,h}}$ and $A_{i-1}\in \mathcal{N}^{A_i}_{A_{i-1}}$ denote the connected user and item nodes respectively, where $\mathcal{N}^{A_i}_{U_{k,h}}$ represents the users connected with $A_i$, and $\mathcal{N}^{A_i}_{A_{i-1}}$ indicates the items connected with $A_i$ in domain A. Then, we can formulate the passed message ($\bm{m}_{A_i \leftarrow U_{k,h}}$) from user $U_{k,h}$ to $A_i$ as:
\begin{equation}
\bm{m}_{A_i \leftarrow U_{k,h}}=\gamma_{U_{k,h}}^{A_i} (\bm{W}_1\bm{e}_{U_{k,h}}+\bm{W}_2(\bm{e}_{U_{k,h}}\odot \bm{e}_{A_i})),
\end{equation}
where $\bm{e}_{U_{k,h}}$ and $\bm{e}_{A_i}$ denotes the presentations of $U_{k,h}$ and $A_i$ respectively. $\gamma_{U_{k,h}}^{A_i}$ is a learnable weight that controls how much information we should pass from $U_{k,h}$ to $A_i$.

Similarly, the passed message from item $A_{i-1}$ to $A_i$ can be defined as:
\begin{align}
\bm{m}_{A_i \leftarrow A_{i-1}}=\gamma_{A_{i-1}}^{A_i} (\bm{W}_1{\bm{e}_{A_{i-1}}}+{\bm{W}_2}(\bm{e}_{A_{i-1}}\odot \bm{e}_{A_i})),
\label{item_passing}
\end{align}
where $\gamma_{A_{i-1}}^{A_i}$ weights the information passed from $A_{i-1}$ to $A_i$. 

Then, we define the retained message of $A_i$ similarly:
\begin{equation}
\bm{m}_{{A_i} \leftarrow {A_i}}=\gamma_{A_i}^{A_i} (\bm{W}_1\bm{e}_{A_i}),
\end{equation}
where $\gamma_{A_i}^{A_i}$ is also a learnable weight (we will introduce $\gamma_{U_{k,h}}^{A_i}$ , $\gamma_{A_{i-1}}^{A_i}$, and $\gamma_{A_i}^{A_i}$ shortly).

However, the above message passing strategy for items (i.e., Eq.~(\ref{item_passing})) ignores the fact that users usually consume different items with different time intervals. For example, users may repeatedly purchase daily consumer products quickly and purchase electronic products in a long time interval. We can also have similar observations in video streaming services, that is, users may watch news programs every day, but only one movie a week.
Intuitively, the time interval between items not only determines users' needs, but also is an important signal for us to distinguish and identify items.

To further enhance the item representation learning, we incorporate the time interval information into the message passing process, and devise a time interval-aware item representation learning method. Let $\bm{t}_{A_{i,i-1}}$ be the representation of the time interval $\delta_{i,i-1} = |t_i -t_{i-1}|$ between items $A_i$ and $A_{i-1}$, where $t_i$ is the timestamp of $A_i$. 
The time intervals between any two adjacent items in the behavior sequence are represented as learnable embeddings, which are initialized by the Xavier method~\cite{glorot2010understanding} and optimized by the Adam algorithm~\cite{kingma2014adam}.
Moreover, all the time interval values (i.e., $\delta_{i,i-1}$) are further clipped into a proper range, since a too large or too small time interval tends to be meaningless.

Then, the passed message from items can be reformulated by adding the time interval with a weighted sum:
\vspace{-0.2cm}
\begin{align}
& \bm{m}_{A_i \leftarrow A_{i-1}}=\gamma_{A_{i-1}}^{A_i} (\bm{W}_1({\alpha\bm{e}_{A_{i-1}}}+(1-\alpha)\bm{t}_{A_{i,i-1}}) \notag\\
& \qquad \qquad \quad +{\bm{W}_2}(\bm{e}_{A_{i-1}}\odot \bm{e}_{A_i})),
\end{align}
where $\alpha$ is a hyper-parameter that controls to what extent we should pay attention to the time interval information. To this end, we have passed three kinds of information to the target item, i.e., the message from $A_{i-1}$, the interacted information between $A_i$ and $A_{i-1}$, and the representation of the corresponding time interval (the message defined in Eq.~(\ref{item_passing}) is the strategy utilized in \ac{DA-GCN}).

\textbf{Sequence-aware Attention Mechanism.} 
To distinguish the importance of the linked users and items, we develop another attention mechanism to measure their sequential depnedencies to the target item, which can be defined as follows:
\begin{align}
& s_{A_{i-1}}^{A_i}=f(\bm{e}_{A_{i-1}},\bm{e}_{A_i}), \\
& s_{U_{k,h}}^{A_i}=f(\bm{e}_{U_{k,h}},\bm{e}_{A_i}),\\
& \quad s_{A_{i}}^{A_i}=f(\bm{e}_{A_{i}},\bm{e}_{A_i}),
\end{align}
where $s_{A_{i-1}}^{A_i}$ and $s_{U_{k,h}}^{A_i}$ denote the importance of item $A_{i-1}$ to $A_i$, and the importance of user $U_{k,h}$ to item $A_i$ respectively. $s_{A_{i}}^{A_i}$ is the importance of the self-connection to $A_i$.
Then, we can reach $\gamma_{A_{i-1}}^{A_i}$ by the following operation: $s_{A_{{i-1}}}^{A_i}$ as:
\begin{equation}
\begin{aligned}
 \gamma_{A_{i-1}}^{A_i} = \text{exp}(s_{A_{i-1}}^{A_i})/ (\sum_{{A_{i-1}'} \in \mathcal{N}_{A_{i-1}}^{A_i}}^{}\text{exp}(s_{{A_{i-1}}^\prime}^{A_i})\\
 { + \sum_{{U_{k,h}}^\prime \in \mathcal{N}_{U_{k,h}}^{A_i}}^{}\text{exp}(s_{{U_{k,h}}^\prime}^{A_i})+s_{A_i}^{A_i}}).
 \end{aligned}
\end{equation}
Similarly, we can obtain $\gamma_{U_{k,h}}^{A_i}$ and $\gamma_{A_i}^{A_i}$ in the same way.

\textbf{Message Aggregation.} 
Then, the representation of the target item $A_i$ is updated by aggregating all the messages from its connected users and items within the same domain:
\begin{align}
& \bm{e}_{A_i} = 
\text{LeakyReLU}(\bm{m}_{{A_i} \leftarrow {A_i}} + \sum_{A_{i-1}\in \mathcal{N}^{A_i}_{A_{i-1}}}{\bm{m}_{A_i \leftarrow A_{i-1}}}
\notag \\
& \qquad
+\sum_{U_{k,h} \in \mathcal{N}^{A_i}_{U_{k,h}}}{\bm{m}_{A_i \xleftarrow{} U_{k,h}}}).
\end{align}

For clarity, we rename $e_{A_i}$ as ${g}^{h}_{A_i}$ to represent the embedding of item $A_i$ w.r.t the $h$-th latent user. Then, the account-level representation ($\bm{g}_{A_i}$) of item $A_i$ can be formulated as:
\begin{equation}
\bm{g}_{A_i} = \frac{1}{H}\sum^{H}_{h=1}\bm{g}^{h}_{A_i},
\end{equation}
where an average operation is applied to all the latent representations.

\subsection{Matrix-form Propagation Rule}
To efficiently learn the representations for all users and items in both domains, we further formulate the propagation rule in a layer-wise matrix form (from layer $l-1$ to $l$), and define it as:
\begin{align}
& \bm{E}_l=\sigma((\alpha(\bm{\mathcal L_1}+\bm{I}) \bm{E}_{l-1}+(1-\alpha)\bm{\mathcal L_2}\bm{T}_{l-1}) \bm{W}_1 \notag \\
& \quad +\bm{\mathcal L_1} \bm{E}_{l-1}\odot \bm{E}_{l-1} \bm{W}_2),
\end{align}
where $\bm{E}_l\in \mathbb{R}^{(p+n+q)\times d}$ denotes the representations for all the users and items in domain A and B, $\bm{T}_l\in \mathbb{R}^{(p+n+q)\times d}$ denotes all the time intervals in both domains, and $\bm{I}$ is an identity matrix. $\mathcal L_1\in \mathbb{R}^{(p+n+q)\times(p+n+q)}$ is the Laplacian matrix of the \ac{CDS} graph, and $\mathcal L_2\in\mathbb{R}^{(p+n+q)\times(p+n+q)}$ is the time interval Laplacian matrix based on $\mathcal{L}_1$. Their definitions are formulated as:
\begin{align}
& \bm{\mathcal L_1}={
\left[ \begin{array}{ccc}
\bm{Y}_{A_i}{}_{A_{i-1}} & \bm{Y}_{A_i}{}_{U_k} & \bm{0}\\
\bm{Y}_{U_k}{}_{A_i}     & \bm{0}       & \bm{Y}_{U_k}{}_{B_j}\\
\bm{0}      & \bm{Y}_{B_j}{}_{U_k} & \bm{Y}_{B_j}{}_{B_{j-1}},
\end{array} 
\right ]}, \\
& \quad \bm{\mathcal L_2}={
\left[ \begin{array}{ccc}
\bm{T}_{A_i}{}_{A_{i-1}} & \bm{0} & \bm{0}\\
\bm{0} & \bm{0} & \bm{0}\\
\bm{0} & \bm{0} & \bm{T}_{B_j}{}_{B_{j-1}},
\end{array} 
\right ]},
\end{align}
where $\bm{Y}_{A_i}{}_{A_{i-1}}\in \mathbb{R}^{p\times p}$ and $\bm{Y}_{A_i}{}_{U_{k}}\in \mathbb{R}^{p\times n}$ are the attention matrices that weigh the importance of the connected items and users to the target item in domain A, respectively;
$\bm{Y}_{U_k}{}_{A_{i}}\in \mathbb{R}^{n\times p}$ represents the weights of the connected items to users in domain A; $\bm{T}_{A_i A_{i-1}}\in \mathbb{R}^{p\times{p}}$ represents the time interval weight matrix between any given item $A_{i-1}$ and $A_i$. Similarly, in domain B we let $\bm{Y}_{B_j}{}_{U_{k}}\in \mathbb{R}^{q\times n}$ and $\bm{Y}_{B_j}{}_{B_{j-1}}\in \mathbb{R}^{q\times q}$ 
be the weight matrices from users and items to the target item respectively.
$\bm{Y}_{U_k}{}_{B_{j}}\in \mathbb{R}^{n\times q}$ denotes the weights from any given item to the target user;
$\bm{T}_{B_j}{}_{B_{j-1}}\in \mathbb{R}^{q\times{q}}$ represents the time interval weights between any two items in domain B.

Note that, we can achieve the propagation rule of the \ac{DA-GCN} method by setting $\alpha=1$, that is, disabling the time interval information in item representation learning.

\subsection{Sequence Representation Learning\label{sequence_embedding}}
After node representation learning, we can get the sequence-level embedding for $S_A$ (or $S_B$) by directly applying the max pooling operation (as \ac{DA-GCN} does) to the item representations within it ($\bm{g}_{A_1}, \bm{g}_{A_2}, ..., \bm{g}_{A_i}, ... $) (or ($\bm{g}_{B_1}, \bm{g}_{B_2}, ..., \bm{g}_{B_j}, ... $)). However, as the above method only leverages one representative item as the sequence-level representation, users' preferences over other items and their interactive characteristics are both ignored. Moreover, though the average pooling strategy can take all the items into consideration, it treats items equally and cannot learn their different contributions to the sequence. In light of this, we first model the interactive features by introducing an account-aware self-attention network, and then devise a vanilla attention network to attentively aggregate all the items.

Specifically, to learn the interactive characteristics among items (take domain A as an example), we first transform each item $\bm{g}_{A_i}$ within the given sequence ($\bm{g}_{A_1}, \bm{g}_{A_2}, ..., \bm{g}_{A_i}, ... $) and account $\bm{e}_{U_k}$ into key $\bm{k}_{A_i} = \bm{g}_{A_i} \bm{W}^k$, value $\bm{v}_{A_i} = \bm{g}_{A_i} \bm{W}^k$, and query $\bm{q}_{A_i} = \bm{e}_{U_k} \bm{W}^q$, where $\boldsymbol{W}^k\in \mathbb{R}^{d\times d_k}$ and $\boldsymbol{W}^q\in \mathbb{R}^{d\times d_k}$ are the projection matrices (we set key equal to value). Then, the dot-product attention between two items $A_i$ and $A_j$ is defined as:
\begin{align}
    \text{Attention}(A_i, A_j) = \text{softmax}( \frac{\boldsymbol{q}_{A_i} \boldsymbol{k}_{A_j}^T}{\sqrt{d_k}}),
\label{eq:dot_attention}
\end{align}
where $d_k$ is the dimension of queries and keys. Note that, we do not follow the standard dot-product attention calculation method~\cite{Vaswani2017attention}, instead, we project the embeddings of accounts as queries to explore their preferences in modeling items interaction mode. On the basis of the obtained weights on values in Eq.~(\ref{eq:dot_attention}), we can further re-define the item representation of $A_i$ as:
\begin{align}
    \bm{g}'_{A_i} = \sum_{j=1}^{|S_A|}\text{softmax}( \frac{\boldsymbol{q}_{A_i} \boldsymbol{k}_{A_j}^T}{\sqrt{d_k}}) \bm{v}_{A_j}.
\end{align}
After that, we feed each updated item embedding into a fully connected feed-forward neural network with two linear transformations:
\begin{align}
    \text{FFN}(\bm{g}'_{A_i}) = max(0, \bm{g}'_{A_i}\boldsymbol{W}_1 +\boldsymbol{b}_1)\boldsymbol{W}_2+\boldsymbol{b}_2,
\end{align}
where $\boldsymbol{b}$ and $\boldsymbol{W}$ are the bias and weight of the neural network. To avoid losing the original input information, a residual connection is applied to $\text{FFN}(\bm{g}'_{A_i})$, followed by layer normalization~\cite{Vaswani2017attention}.
To explore items' interactive features deeply, we circularly stack the above sub-layers (in experiments, we set the number of layers as 2).
Besides, to make use of the sequence order, following the method in \cite{Vaswani2017attention}, positional encodings are also added to the input embeddings at the bottom of the stack.

To model the different contributions of items to the sequence ($S_A$ or $S_B$), we further devise another attention network to attentively aggregate the items within it (as shown in Fig.~\ref{fig:SRL}). What is more, as users' sequential preferences are often reflected by their recent interactive behaviors, we leverage the last item ($A_t = A_{i-2}$ and $B_t = B_{j-2}$) within the sequence as the supervision signal to guide the sequence representation learning, that is, an attention network is developed to estimate the attention weights of items within the given sequence (we take domain A as an example). 

For the $i$-th item in $S_A$ ($A_i$), we compute its relevance to $A_t$ by:
\begin{align}
a_{i,t} = f(\bm{z}_{A_i}, \bm{z}_{A_t}),
\end{align}
where $\bm{z}_{A_i}$ is the representation of $A_i$ with learned interactive features; $f(\cdot)$ is a score function that measures the correlation between $\bm{z}_{A_i}$ and $\bm{z}_{A_t}$. Inspired by \cite{he2018nais}, we parameterize $f(\cdot)$ by a \ac{MLP} network: 
\begin{align}
f(\bm{z}_{A_i}, \bm{z}_{A_t}) = \bm{W}_2^TReLU(\bm{W}_1[\bm{z}_{A_i}\oplus \bm{z}_{A_t}]+\bm{b})
\end{align}
where $\bm{W}_1$ and $\bm{b}$ are weight matrix and bias vector, respectively, to map the input into a hidden layer. $\bm{W}_2$ is a weight vector that projects the output of the first layer into an attention weight.

Then, we can get the sequence representation with the following attention network:
\begin{align}
& \qquad \quad\bm{h}_{S_A} = \sum_{i=1}^{|S_A|} a_{i,t}\bm{z}_{A_i}, \\
& a_{i,t} = \frac{\text{exp}(f(\bm{z}_{A_i}, \bm{z}_{A_t}))}{[\sum_{i=1}^{|S_A|}\text{exp}(f(\bm{z}_{A_i}, \bm{z}_{A_t}))]^\beta},
\end{align}
where $\bm{h}_{S_A}$ denotes the final representation of sequence $S_A$, which is learned by attentively aggregating the items within it. $\beta \in [0,1]$ is the smoothing exponent~\cite{he2018nais} that aims at suppressing the value of the denominator to avoid overly punishing the attention weights for active users (i.e., sequences that have relatively more items). In experiments, we set $\beta =0.5$.

\subsection{The Prediction Layer\label{prediction_layer}}
After the sequence representation layer, we can get the sequence embedding $\bm{h}_{S_A}$ (or $\bm{h}_{S_B}$) for $S_A$ (or $S_B$). Then, to leverage the information in both domains, we feed the concatenation of $\bm{h}_{S_A}$ and $\bm{h}_{S_B}$ to the prediction layer:
\begin{align}
P(A_{i+1}|S_A, S_B) = softmax(\bm{W}_A \cdot [\bm{h}_{S_A},\bm{h}_{S_B}]^\mathrm{T}+\bm{b}_A),\notag\\
P(B_{j+1}|S_B, S_A) = softmax(\bm{W}_B \cdot [\bm{h}_{S_B},\bm{h}_{S_A}]^\mathrm{T}+\bm{b}_B),\notag
\end{align}
where $\bm{W}_A$ and $\bm{W}_B$ are the embedding matrices of all the items in domain A and B, respectively; $\bm{b}_A$ and $\bm{b}_B$ are the bias terms. Then, we leverage the cross-entropy as our loss function to train \ac{TiDA-GCN} in both domains:
\begin{align}
L_A(\theta ) = -\frac{1}{|\mathbb{S}|}\sum_{S_A, S_B \in \mathbb{S}}\sum_{A_i \in S_A}\text{log} P(A_{i+1}|S_A,S_B), \\
L_B(\theta ) = -\frac{1}{|\mathbb{S}|}\sum_{S_B, S_A \in \mathbb{S}}\sum_{B_j \in S_B}\text{log}P(B_{j+1}|S_A,S_B),
\end{align}
where $\mathbb{S}$ represents the training sequences in both domain A and B, and $\theta$ denotes the model parameters of \ac{TiDA-GCN}. 
To enhance the cross-domain recommendation, a joint training scheme is applied to both domains:
\begin{align}
L(\theta ) = L_A(\theta )+L_B(\theta ).
\end{align}
All the parameters in \ac{TiDA-GCN} are learned via the gradient back-propagation algorithm in an end-to-end training manner.
\begin{table}
    \centering
    \caption{The statistics of our HVIDEO and HAMAZON datasets.}
    \begin{tabular}{lcccccc}
    \toprule
    & \multicolumn{3}{c}{HVIDEO}&\multicolumn{3}{c}{HAMAZON} \\
    \cmidrule{1-7}
    \multirow{1}[1]{*}{}& \multicolumn{3}{c}{E-domain} & \multicolumn{3}{c}{M-domain}\\
    \#Items &\multicolumn{3}{c}{8,367} &\multicolumn{3}{c}{67,161}\\
    \#Interactions &\multicolumn{3}{c}{2,129,500} &\multicolumn{3}{c}{2,196,574}\\
    \#Avg. sequence length & \multicolumn{3}{c}{15.85}
    & \multicolumn{3}{c}{15.27} \\
    \cmidrule{1-7}
    \multirow{1}[1]{*}{}& \multicolumn{3}{c}{V-domain} & \multicolumn{3}{c}{B-domain}\\
    \#Items &\multicolumn{3}{c}{11,404} &\multicolumn{3}{c}{126,547}\\
    \#Interactions &\multicolumn{3}{c}{1,893,784} &\multicolumn{3}{c}{2,135,995}\\
    \#Avg. sequence length & \multicolumn{3}{c}{14.09}
    & \multicolumn{3}{c}{14.84} \\
    \cmidrule{1-7}
    \#Accounts & \multicolumn{3}{c}{13,714}
    & \multicolumn{3}{c}{13,724} \\
    \#Sequences & \multicolumn{3}{c}{134,349}
    & \multicolumn{3}{c}{143,885} \\
    \#Train-sequences & \multicolumn{3}{c}{114,197}
    & \multicolumn{3}{c}{122,303} \\
    \#Test-sequences & \multicolumn{3}{c}{20,152}
    & \multicolumn{3}{c}{21,582} \\
    \bottomrule
    \end{tabular}
    \label{tab:dataset_statistics}
\end{table}

\section{Experimental Setup\label{Experimental_setup}}
This section first introduces the research questions to be answered in experiments, and then describes the datasets, evaluation protocols, and baselines utilized in our work.

\subsection{Research Questions} 
We intend to answer the following research questions: 
\begin{itemize}
    \item[\textbf{RQ1}] How does our devised \ac{TiDA-GCN} method perform compared with other state-of-the-art methods?
    \item[\textbf{RQ2}] What is the performance of \ac{TiDA-GCN} on different domains? Can we benefit from the cross-domain information?
    \item[\textbf{RQ3}]
    Is it helpful to leverage the sequence and time interval information in learning node representations? How do the key components of \ac{TiDA-GCN}, i.e., attention mechanism, time interval modeling, and sequence modeling, contribute to the recommendation performance?
    \item[\textbf{RQ4}] How do the hyper-parameters (i.e., $H$ and $\alpha$) affect the performance of our solution?
    \item[\textbf{RQ5}] How is the training efficiency of our \ac{TiDA-GCN} method?
\end{itemize}

\begin{table*}
  \centering
  \caption{Comparison results on HVIDEO and HAMAZON. Note that, VUI-KNN does not work on HAMAZON dataset because it needs specific time in a day to split the virtual users, which is not available in this dataset.}
  \normalsize 
  \resizebox{\textwidth}{4.5cm}{
	\begin{threeparttable} 
    \begin{tabular}{lcccccccc|cccccccc}
    \toprule
    \multicolumn{1}{c}{\multirow{4}[4]{*}{\textbf{Methods}}} & \multicolumn{8}{c|}{\textbf{HVIDEO}} &
    \multicolumn{8}{c}{\textbf{HAMAZON }} \\
    \cmidrule{2-17}
    & \multicolumn{4}{c}{\textbf{E-domain (\%) }} & \multicolumn{4}{c|}{\textbf{V-domain (\%) }} & \multicolumn{4}{c}{\textbf{M-domain (\%)}} & \multicolumn{4}{c}{\textbf{B-domain (\%)}} \\
    \cmidrule{2-17}
          & \multicolumn{2}{c}{MRR} & \multicolumn{2}{c}{Recall} 
          & \multicolumn{2}{c}{MRR} & \multicolumn{2}{c|}{Recall}
          &
          \multicolumn{2}{c}{MRR} & \multicolumn{2}{c}{Recall}
          & \multicolumn{2}{c}{MRR} & \multicolumn{2}{c}{Recall} \\
    \cmidrule{2-17}
          & @5      & @20   & @5    & @20   & @5    & @20   & @5    & @20  & @5     & @20   & @5    & @20   & @5    & @20   & @5    & @20
          \\
    \midrule
    Item-KNN &{3.52} &{4.20} &{3.73} &{9.41} &{5.14} &{12.97} &{8.85} &{13.21} 
    &{1.95} &{2.22} &{5.55} &{6.77} &{4.22} &{5.81} &{6.62} &{9.33} \\
    BPR-MF &{2.82} &{3.33} &{2.52} &{5.56} &{4.12} &{7.63} &{5.26} &{9.32} 
    &{2.04} &{2.35} &{4.21} &{5.52} &{2.48} &{3.92} &{1.57} &{4.41} \\
    NCF &{5.77} &{7.85} &{11.25} &{20.12} &{22.99} &{24.71} &{27.21} &{22.63} 
    &{2.72} &{3.11} &{7.82} &{10.34} &{7.92} &{9.88} &{8.31} &{11.22} \\
    LightGCN &{11.55} &{13.56} &{20.70} &{39.92} &{54.00} &{56.72} &{58.45} &{63.55} 
    &{12.88} &{13.12} &{15.54} &{18.20} &{15.58} &{17.30} &{21.14} &{22.88} \\
    \midrule
    VUI-KNN &{3.21} &{4.08} &{3.44} &{6.92} &{4.32} &{8.30} &{7.85} &{9.08} 
    &\textbf{-} &\textbf{-} &\textbf{-} &\textbf{-} &\textbf{-} &\textbf{-} &\textbf{-} &\textbf{-} \\
    \midrule
    NCF-MLP++ &{6.55} &{8.82} &{11.37} &{21.92} &{26.67} &{29.15} &{31.70} &{37.90} 
    &{5.55} &{6.23} &{13.32} &{14.77} &{10.29} &{12.60} &{14.01} &{14.77} \\
    Conet &{7.88} &{9.52} &{13.30} &{25.52} &{28.24} &{30.11} &{34.76} &{38.21} 
    &{6.51} &{6.72} &{13.45} &{15.21} &{10.43} &{12.77} &{14.11} &{15.02} \\
    \midrule
    GRU4REC &{13.33} &{15.48} &{22.47} &{44.18} &{61.80} &{62.47} &{66.59} &{73.23} 
    &{13.76} &{14.16} &{16.88} &{18.86} &{16.88} &{17.18} &{20.89} &{23.64} \\
    HGRU4REC &{13.76} &{16.14} &{22.55} &{47.98} &{60.58} &{63.31} &{68.00} &{73.24} 
    &{13.75} &{14.14} &{16.96} &{20.81} &{17.04} &{17.35} &{20.92} &{23.64} \\
    NAIS &11.45&13.24&19.80&40.17&51.52&54.47&59.30&67.41
    &10.55&12.57&14.03&16.02&13.29&15.99&14.51&19.82\\
    \midrule
    Time-LSTM &11.19&13.27&20.17&41.45&59.73&60.42&64.66&71.59
    &11.16&11.60&13.91&18.35&13.74&14.13&17.35&21.17\\
    TGSRec &13.95&15.73&19.91&41.80&50.32&53.40&58.91&67.22
    &11.91&14.00&14.59&18.44&15.74&16.63&14.57&19.93\\
    \midrule
    $\pi$-Net &{15.36} &{17.52} &{25.13} &{47.08} &{60.37} &{61.74} &{67.00} &{74.17} 
    &{16.24} &{16.56} &{18.54} &{21.87} &{20.38} &{20.58} &{22.44} &{23.79} \\
    PSJNet &{15.37} &{17.56} &{24.80} &{46.68} &{61.89} &{62.63} &{66.86} &{74.14} 
    &11.25 &13.58 &16.25 &18.14 &15.52 &17.30 &16.67 &19.30 \\
    \midrule
    DA-GCN &35.63&37.27&51.35&66.93&59.78&60.55&75.39&82.37
    &20.09&20.19&22.93&23.90&21.35&21.39&23.93&24.25\\
    TiDA-GCN &\textbf{38.66}$^\dag$ &\textbf{40.23}$^\dag$ &\textbf{54.11}$^\dag$ &\textbf{68.98}$^\dag$ &\textbf{63.58}$^\dag$ &\textbf{65.37}$^\dag$ &\textbf{76.37}$^\dag$ &\textbf{83.58}$^\dag$ 
    &\textbf{20.91}$^\dag$&\textbf{21.23}$^\dag$&\textbf{23.55}$^\dag$&\textbf{24.33}&\textbf{21.88}&\textbf{22.21}$^\dag$&\textbf{24.69}$^\dag$&\textbf{24.82}\\
    \bottomrule
    \end{tabular}%
\begin{tablenotes} 
    \normalsize
    \item \textbf{Bold value} represents the best result of the compared methods in terms of the corresponding metric. Significant improvements over the best baseline results (DA-GCN) are marked with $^\dag$ (t-test, $p <$ .05). 
\end{tablenotes}
\end{threeparttable}}
  \label{tab:resutls}
\end{table*}

\subsection{Datasets} 
We evaluate our \ac{TiDA-GCN} method on two real-world datasets that are released for the \ac{SCSR} task, i.e., HVIDEO~\cite{ma2019pi} and HAMAZON\cite{ren2019net}.

HVIDEO is collected from a smart TV platform that involves the watching logs of family accounts on two domains, i.e., the education domain (E-domain) and video domain (V-domain), from Oct. 2016 to Jun. 2017. The E-domain refers to the educational videos for students at all stages, and instructional videos on daily life. The V-domain refers to the videos from the video-on-demand platform 
on movies, cartoons, and TV series, etc.
As family accounts are usually shared by family members and their watching logs are the mixed behaviors of a family, this dataset is suitable for the \ac{SCSR} task.
To reduce the impact of noisy data, we further filter out the accounts with less than 10 watching records and those logs with less than 300 seconds of watching time. As a result, our final data keeps 13,714 family accounts and 134,349 watching sequences.

HAMAZON is another benchmark dataset for \ac{SCSR}, which contains the user review records on two Amazon domains, i.e., movie domain (M-domain) and book domain (B-domain), from May 1996 to July 2014. The M-domain specifies the watching and reviewing logs from Amazon users to movies. The B-domain refers to the reading and rating behaviors of Amazon users on books. As the original Amazon data is not released for \ac{SCSR}, the domain and account information cannot directly meet our requirement. Therefore, we follow the strategy in \cite{ren2019net}, and preprocess the Amazon data by the following steps: 1) As in our settings, users can simultaneously interact with items from different domains; we only retain the users who have interactions in both domains. 2) To simulate the shared-account characteristic, we first split the schedule into six intervals (i.e., 1996-2000, 2001-2003, 2004-2006, 2007-2009, 2010-2012, 2013-2015). Then, in each interval, we randomly merge 2-4 users and their interaction records as one shared account. After that, we split each sequence within one year into small fragments, and further filter out all the sequences with less than five items in M-domain and two items in B-domain to achieve our final data. The resulting data contains 13,724 accounts and 143,885 sequences. The statistics of HVIDEO and HAMAZON are reported in Table~\ref{tab:dataset_statistics}.

Note that, due to missing the time interval information in our previously prepossessed data~\cite{guo_DAGCN_2021}, they cannot be directly applied to our solutions. Hence, we regenerate both datasets from their original form, and re-conduct all the experiments accordingly.

\subsection{Evaluation Protocols}
In the experiments, we randomly select 85\% of all the sequences as the training data and the remaining as the test set for both HVIDEO and HAMAZON. For evaluation, we treat the last item in each sequence for each domain as the ground truth, and intend to evaluate the ability of our model in predicting the next item given one account's historical behavior sequence.

We explore the effectiveness of our proposed models by reporting the top-$N$ recommendation results measured by two commonly used metrics, Recall@$N$ and MRR@$N$~\cite{hidasi2016srnn,Guo_streaming_2019} (we set $N=\{5, 20\}$ in the experiments).
\begin{itemize}
    \item Recall@$N$ measures how many desired items of the test data can be covered amongst the top-$N$ recommended items. Note that Recall does not consider the actual rank of the item, and it can work well in practical scenarios where the absolute item order does not matter.
    \item MRR@$N$ (Mean Reciprocal Rank) computes the average of the reciprocal ranks of the desired items, and the reciprocal rank above $N$ is set to zero. Compared with Recall, MRR takes into account the rank order of the item, and it is an important measurement for the recommendation scenarios where the item ranking order matters.
\end{itemize}

\subsection{Baseline Methods}
We evaluate \ac{TiDA-GCN} by comparing it with the following baselines in six categories: traditional, shared-account, cross-domain, sequential, time interval-aware sequential, and shared-account cross-domain sequential recommendations.

1) Traditional recommendation methods.
For comparison, we adapt them to the \ac{SR} task and report their results in each domain.

\begin{itemize}
    \item Item-KNN~\cite{greg2003knn}: 
    This is a simple yet frequently used baseline that recommends items according to the item-to-item similarity, defined as the cosine similarity of two items measured by the sequences in which they appear.
    \item BPR-MF~\cite{hidasi2016srnn}: This is a matrix factorization method that is commonly used as a ranking-based baseline. We apply it to \ac{SR} by representing sequences with the average latent factors of items appearing in them.
    \item NCF~\cite{he2017neural}: 
    This is a collaborative filtering method implemented by a neural network. We exploit it for \ac{SR} by treating users as sequences and learning their representations via an average pooling layer over the latent factors of items within them.
    \item LightGCN~\cite{He2020lightGCN}:  This method models the sequences and collaborative filtering mechanism by a graph convolutional network, where an item representation is learned by aggregating the information from its neighbors.
\end{itemize}

2) Shared-account recommendation methods. These kinds of methods are devised for making recommendations for shared accounts~\cite{li2009can,verstrepen2015top,yang2017personalized}. However, most of them need extra information unavailable in our dataset, such as explicit ratings and item descriptions, for user identification. Hence, we select the following method that needs similar input to ours as comparisons. 

\begin{itemize}
    \item VUI-KNN~\cite{wang2014user}:
    This method supposes different users within a shared account should get used to consuming services during different periods. Then, it decomposes users in a composite account by dividing consuming logs into periods, which are further viewed as virtual users. The users in an account are identified by the combinations of virtual users.
\end{itemize}
 3) Cross-domain recommendation methods. 
\begin{itemize}
    \item NCF-MLP++~\cite{ma2019pi}:
    This is an improved NCF method~\cite{he2017neural} that shares the collaborative filtering in different domains.
    \item Conet~\cite{hu2018conet}: 
    This is a cross-domain recommendation method on the basis of a cross-stitch network, where the domain sharing mechanism is modeled by a neural collaborative filtering model.
\end{itemize}

4) Sequential recommendation methods. As these methods are mainly proposed for the sequential recommendation on a single domain, we report their results on each domain.
\begin{itemize}
    \item GRU4REC~\cite{hidasi2016srnn}: This is an RNN-based sequential recommendation method, which employs a GRU component and a ranking-based loss function to learn users' sequential patterns.
    \item HGRU4REC~\cite{quadrana2017hrnn}: This method further improves GRU4REC by incorporating a hierarchical RNN structure, where the sequential relations among sessions and user identities are simultaneously considered.
    \item NAIS~\cite{he2018nais}: This is an item-based CF method that explores a nonlinear attention network to learn item similarities. The neural attention design enables NAIS to distinguish the importance of items in users' historical sequential behaviors. 
\end{itemize}

5) Time interval-aware sequential recommendation methods.
\begin{itemize}
    \item Time-LSTM~\cite{ZhuLLWGLC17}: This is a time interval-aware recommendation method that models time gaps between users' actions with time gates in a new LSTM variant.
    \item TGSRec~\cite{FanLZX0Y21}: This is another time interval-aware sequential recommendation method that explicitly models collaborative
    signals and time gaps among items via a collaborative attention network, where the time span is represented as a kernel value of the corresponding encoded time embeddings.
\end{itemize}

6) Shared-account cross-domain sequential recommendation methods.
\begin{itemize}
    \item $\pi$-Net~\cite{ma2019pi}: This is a state-of-the-art method proposed for \ac{SCSR}, which explores a parallel RNN and a gating mechanism to transfer the shared information across domains. For the shared-account characteristic, a Shared account Filter Unit (SFU) is applied to learn user-specific representations.
    
    \item PSJNet~\cite{ren2019net}: This is another recently developed method for \ac{SCSR}, which improves $\pi$-Net via splitting the mixed representations to get role-specific representations and joining them to get cross-domain representations. Note that, due to the limitation of our GPU memory, we fail to get the expected results on the HAMAZON dataset. To complete the training process on HAMAZON, we only select 80,000 sequences to optimize PSJNet.
    
    \item DA-GCN~\cite{guo_DAGCN_2021}: This is a preliminary version of our work, which achieves the state-of-the-art performance in our previous experiments.
    However, this method does not consider the time interval information between items, which is incapable of distinguishing the timeliness of sequential correlations between them. Moreover, DA-GCN fails to consider the importance of different items and their interactive characteristics, hindering us from learning more accurate representations.
\end{itemize}

\subsection{Implementation Details}
We implement our proposal based on the Tensorflow platform, and accelerate the training processes using a Quadro RTX 6000 GPU.
For model optimization, we use the Xavier method \cite{glorot2010understanding} to randomly initialize all the parameters, and take Adam~\cite{kingma2014adam} as our optimization algorithm.
For the hyper-parameters of the optimizer, we set the learning rate as 0.001, and the batch size as 256. The dropout ratio is set as 0.1 at the model training stage on both datasets. For the model hyper-parameters, we set the embedding size of both users and items as 16. 
For the number of users sharing an account ($H$), we search its value within $\{1, 2, 3, 4, 5\}$.
For the importance of the time interval information ($\alpha$), we search its value in [0, 1] with step size 0.1.

For the hyper-parameters in baselines, we refer them to the settings in their publications and fine-tune them on different datasets. For example:
1) For Item-KNN, we set the neighbor size per item as 20. For BPR-MF, we set the embedding size of both users and items as 16. 2) For NCF, NCF-MLP++, and NAIS, we enhance the training data by augmenting four negative samples per positive.
3) For VUI-KNN, we follow its strategy by cutting a day into six pieces to simulate the shared-account characteristic. 
4) For Light-GCN, we set its layer number as 4, and the embedding size of its hidden unit as 16. 5) For Conet, we set the class size and the input size both as 2 to make it comparable with ours. 6) For GRU4REC and HGRU4REC, we set the learning rates as 0.005 and 0.001, respectively. The embedding sizes of their hidden GRU units are both set as 16. 7) For $\pi$-Net and PSJNet, we set the number of latent users sharing an account both as 4 to reach their best performance. 8) For TGSRec, we set the dimension of node and time interval representations both as 16, and the number of heads as 2. 9) For Time-LSTM, we set the number of its hidden units as 128.

Note that, for a fair comparison, we do not follow the evaluation methods in NCF, NCF-MLP++, and NAIS, which only test their ability in ranking 100 randomly sampled negatives per positive; instead, we treat all the items as the candidate set to ensure the stability of the methods. Moreover, we conduct the one sample paired t-tests to verify the significance of our improvement over the baseline results ($p< .05$).

\begin{table*}
  \centering
  \caption{Ablation studies on HVIDEO and HAMAZON.}
  \scriptsize
    \begin{tabular}{lcccccccc|cccccccc}
    \toprule
    \multicolumn{1}{c}{\multirow{4}[2]{*}{\textbf{Variants}}} &
    \multicolumn{8}{c|}{\textbf{HVIDEO}} &
    \multicolumn{8}{c}{\textbf{HAMAZON}}\\
    \cmidrule{2-17}
    & \multicolumn{4}{c}{\textbf{E-domain (\%)}} &
    \multicolumn{4}{c|}{\textbf{V-domain (\%)}} &
    \multicolumn{4}{c}{\textbf{M-domain (\%)}} &
    \multicolumn{4}{c}{\textbf{B-domain (\%)}}\\
    \cmidrule{2-17}
          & \multicolumn{2}{c}{MRR} & \multicolumn{2}{c}{Recall} 
          & \multicolumn{2}{c}{MRR} & \multicolumn{2}{c|}{Recall}
          & \multicolumn{2}{c}{MRR} & \multicolumn{2}{c}{Recall} 
          & \multicolumn{2}{c}{MRR} & \multicolumn{2}{c}{Recall}
          \\
    \cmidrule{2-17}
          & @5 & @20 & @5 & \multicolumn{1}{c}{@20}
          & @5 & @20 & @5 & \multicolumn{1}{c|}{@20} 
          & @5 & @20 & @5 & \multicolumn{1}{c}{@20}
          & @5 & @20 & @5 & @20 
          \\
    \midrule
    GCN-S &35.52&37.24&{51.07}&{66.41}&{59.48}&{60.22}&{75.28}&{82.12}
    &{19.22}&{19.24}&{20.78}&{21.22}&{21.03}&{21.05}&{23.43}&{23.59}\\
    GCN-A &{35.14}&{36.91}&{50.81}&{65.99}&{59.13}&{60.00}&{74.99}&{81.58}
    &{18.57}&{19.05}&{18.99}&{19.73}&{19.91}&{20.05}&{21.34}&{23.0}\\
    GCN-AS &{34.94}&{36.80}&{50.74}&{65.55}&{59.07}&{59.90}&{74.77}&{81.23}
    &{18.42}&{18.63}&{20.71}&{21.50}&{19.22}&{19.49}&{22.01}&{22.80}\\
    \midrule
    TGCN-Ti &36.38&38.05&51.61&67.71&60.55&61.39&75.57&83.30
    &{20.58}&{20.66}&{23.09}&{23.83}&{21.09}&{21.13}&{23.86}&{24.22}\\
    TGCN-M &35.38&37.34&50.81&66.61&59.57&60.36&74.97&82.36
    &{18.77}&{19.03}&{21.54}&{22.90}&{20.13}&{20.42}&{22.87}&{23.48}\\
    \midrule
    \ac{DA-GCN} &35.63&37.27&51.35&66.93&59.78&60.55&75.39&82.37
    &20.09&20.19&22.93&23.90&21.35&21.39&23.93&24.25\\
    TiDA-GCN &\textbf{38.66} &\textbf{40.23} &\textbf{54.11} &\textbf{68.98} &\textbf{63.58} &\textbf{65.37} &\textbf{76.37} &\textbf{83.58} 
    &\textbf{20.91}&\textbf{21.23}&\textbf{23.55}&\textbf{24.33}&\textbf{21.88}&\textbf{22.21}&\textbf{24.69}&\textbf{24.82}\\
    \bottomrule
    \end{tabular}%
  
  \label{tab:ablation}%
\end{table*}%
\section{Experimental Results (RQ1 \& RQ2)\label{results}}
The comparison results with baselines on HVIDEO and HAMAZON are presented in Table~\ref{tab:resutls}, from which we have the following observations:
1) \ac{TiDA-GCN} and \ac{DA-GCN} outperform other baselines on both datasets, demonstrating the effectivenesses of our proposals in solving the \ac{SCSR} task, that is, the capability of our models in capturing the shared-account characteristics, structural cross-domain knowledge, time interval information, and users' sequential behaviors. Moreover, \ac{TiDA-GCN} and \ac{DA-GCN} achieve the best performance on both domains, indicating the benefits of considering the cross-domain information, and the capabilities of our solutions in transferring the domain knowledge.
2) \ac{TiDA-GCN} outperforms \ac{DA-GCN}, denoting the importance of considering the time interval information and modeling the sequence representations from a multi-aspect view. \ac{TiDA-GCN} and \ac{DA-GCN} perform better than RNN-based methods (i.e., GRU4REC, HGRU4REC, $\pi$-Net, and PSJNet), showing the advantage of leveraging the \ac{GCN} technique to model the multiple relationships among users and items. Furthermore, \ac{TiDA-GCN} and \ac{DA-GCN} have a better performance than $\pi$-Net and PSJNet, again demonstrating the importance of the structural cross-domain and time interval information, and the advantage of our proposals in modeling the cross-domain characteristic.
3) Our GCN-based methods (\ac{TiDA-GCN} and \ac{DA-GCN}) achieve a better performance than the sequential recommendation methods (i.e., GRU4REC, HGRU4RE, NAIS, Time-LSTM, and TGSRec), indicating the importance of the cross-domain knowledge in making recommendations for users with mixed behaviors. The gap between \ac{TiDA-GCN} and traditional recommendation methods (e.g., BPR-MF, NCF, and Light-GCN) shows the importance of the shared-account and cross-domain characteristics, and the effectiveness of our solutions.
4) The improvement of our proposed methods (i.e., \ac{TiDA-GCN} and \ac{DA-GCN}) over the methods that do not consider the shared-account information (e.g., NAIS, Conet, and Light-GCN), denoting the benefit of modeling an account from a multi-user view. From Table~\ref{tab:resutls}, we also notice that only modeling the shared-account characteristic (e.g., VUI-KNN) while ignoring the sequential patterns of user behaviors, cannot get better results than considering them simultaneously.
\section{Experimental Analysis\label{analysis}}
In this section, we first conduct ablation studies to analyze the importance of different model components. Then, we investigate the impact of hyper-parameters and the training efficiency of our solutions.

\begin{figure*}
    \centering
    \includegraphics[width=18cm]{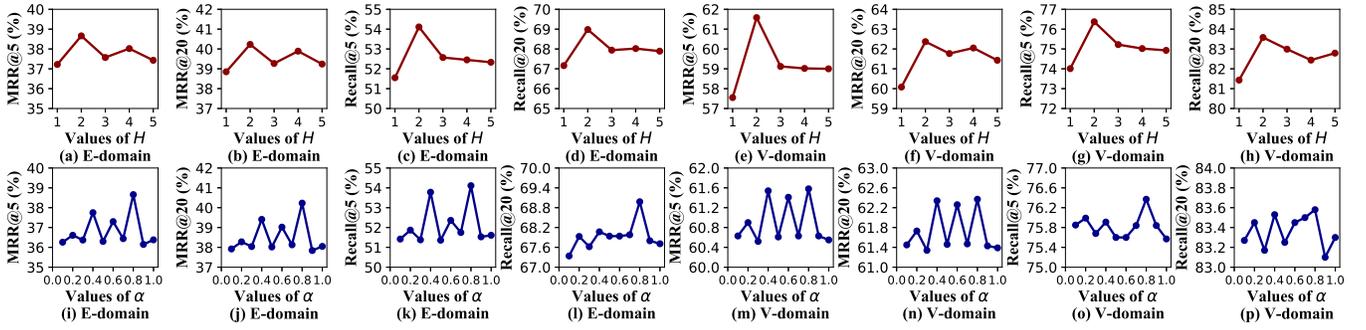}
    \caption{Impact of the hyper-parameters $H$ and $\alpha$ to \ac{TiDA-GCN} on HVIDEO.}
    \label{fig:hyper_account}
\end{figure*}

\subsection{Ablation Studies (RQ3)}

To explore the importance of different components in \ac{TiDA-GCN} and \ac{DA-GCN}, we conduct ablation studies on both datasets while keeping all the hyper-parameters fixed. 
We compare our solutions with the following variants of them:
\begin{itemize}
    \item GCN-S: This is a variant of \ac{DA-GCN} that disables the sequential information when constructing the CDS graph, that is, ignoring the sequential relationships among the items in both domains.
    \item GCN-A: This is another variant of \ac{DA-GCN} that does not exploit the attention mechanisms when aggregating the passed messages.
    \item GCN-AS: This variant of \ac{DA-GCN} removes both components (i.e., the attention mechanisms and the sequential information) from it. This is to demonstrate the benefits brought by considering the sequential information and attention mechanism for \ac{SCSR}.
    \item TGCN-Ti: This is a variant of \ac{TiDA-GCN} that disables the time interval information when learning user and item representations. We conduct this experiment to investigate the importance of the time interval information in the graph neural network.
    \item TGCN-M: This method removes the sequence learning component from \ac{TiDA-GCN}, and only utilizes a simple max-pooling operation to learn the account representations from the sequences (as DA-GCN does). 
\end{itemize}
Note that, \ac{DA-GCN} can also be viewed as a variant of \ac{TiDA-GCN} that removes both the time interval and sequence learning module from it. 

The experimental results are shown in Table~\ref{tab:ablation}, from which we have the following observations:
1) \ac{DA-GCN} outperforms GCN-S and GCN-AS on both datasets, demonstrating the importance of modeling the sequential relationships among items, and the effectiveness of our GCN-based solution.
2) \ac{DA-GCN} achieves a better result than GCN-A and GCN-AS, showing the importance of the attention mechanism in aggregating the passed messages. That is, treating each neighbor differently can result in better node representations than weighting them equally. The gap between \ac{DA-GCN} and GCN-AS again demonstrates the importance of the sequential information and the attention mechanism.
3) \ac{TiDA-GCN} performs better than TGCN-Ti, denoting the significance of incorporating the time interval information into the node representation learning process. From this result, we can indicate that the time intervals among items provide us evidence to distinguish different types of item pairs, which leads to a better recommendation result.
4) \ac{TiDA-GCN} has a better performance than TGCN-M, demonstrating the importance of our sequence learning module. That is, learning items' interactive characteristics and treating them differently can benefit the sequence-level account representation. The improvement of \ac{TiDA-GCN} over \ac{DA-GCN} shows the benefits brought by our time interval and sequence modeling components.

\begin{figure}
    \centering
    \includegraphics[width=8.5cm]{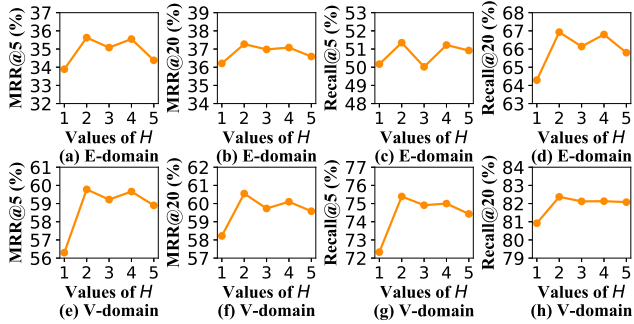}
    \caption{Impact of the hyper-parameter $H$ to \ac{DA-GCN} on HVIDEO.}
    \label{fig:da_param}
\end{figure}

\subsection{Impact of Hyper-parameters (RQ4).}
In this section, we further investigate the impact of hyper-parameters to \ac{DA-GCN} and \ac{TiDA-GCN} on HVIDEO (similar trends are obtained on HAMAZON), and report the experimental results in Fig.~\ref{fig:hyper_account} and Fig.~\ref{fig:da_param}.

\textbf{Impact of hyper-parameter $H$:} 
$H$ is introduced as a hyper-parameter in both \ac{DA-GCN} and \ac{TiDA-GCN} to denote the number of latent users sharing an account. Considering the common size of shared accounts in reality, we search its value within \{1, 2, \dots 5\}. The experimental results are shown in Fig.~\ref{fig:hyper_account} and Fig.~\ref{fig:da_param}, from which find that both \ac{DA-GCN} and \ac{TiDA-GCN} perform worst when setting $H=1$, and perform best when $H=2$. Moreover, we also observe that both of them are getting worse with further increasing the value of $H$. This result demonstrates that viewing an account as a virtual user cannot get a better result than modeling it as $H$ latent users, which is more conducive to discovering the interests of different users from mixed behaviors.
    
\textbf{Impact of hyper-parameter $\alpha$:} To further explore the impact of considering time interval in generating the node representations, we introduce the hyper-parameter $1-\alpha$ within [0, 1] to denote its contribution to \ac{TiDA-GCN}. When setting $1-\alpha=0$, meaning that we will disable the time interval formation in the graph convolution process, and vice visa. From the experimental results presented in Fig.~\ref{fig:hyper_account}, we can observe that the time interval information impacts \ac{TiDA-GCN} significantly, and it reaches its best result when $\alpha=0.8$ on most metrics. From the results, we also find that just disabling or relying too much on the time interval information cannot get better performance than combining it with the node information probably.

\subsection{Training Efficiency and Scalability (RQ5).}

\begin{figure}
    \centering
    \includegraphics[width=8.5cm]{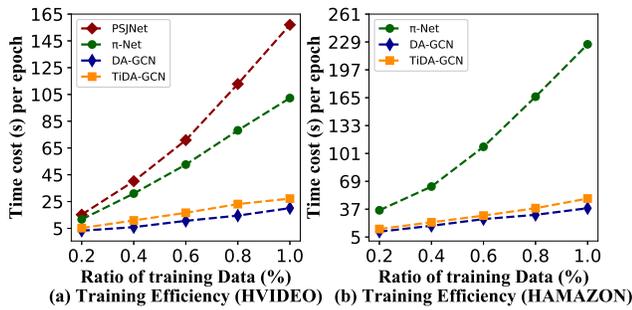}
    \caption{Model Training Efficiency on HVIDEO and HAMAZON. Due to the GPU memory size, we cannot run PSJNet on HAMAZON with its full data.}
    \label{fig:training_efficiency}
\end{figure}

To explore the training efficiency and scalability of \ac{DA-GCN} and \ac{TiDA-GCN}, we further investigate their time cost of model training with different dataset proportions (i.e., \{0.2, 0.4, 0.6, 0.8, 1.0\}). The comparison results with other state-of-the-art recommendation methods for \ac{SCSR} are reported in Fig.~\ref{fig:training_efficiency}. From these results, we can find that: 
1) Our \ac{DA-GCN} and \ac{TiDA-GCN} methods need less training time than other baselines on both datasets, demonstrating the higher training efficiency of our models in dealing with mixed user behaviors. The reason that training \ac{TiDA-GCN} needs to pay more time than \ac{DA-GCN} is mainly because \ac{TiDA-GCN} further considers the time interval and interactive characteristics of items in the representation learning process.
2) From the results, we also observe that with the dataset ratio gradually increasing from 0.2 to 1.0, the time costs for training \ac{DA-GCN} and \ac{TiDA-GCN} grow almost linearly, which demonstrates the scalability of our model to large-scale data and provides us positive evidence to answer RQ5.

\section{Conclusion and Future Work}
In this work, we propose \ac{TiDA-GCN} for \ac{SCSR} to learn user and item representations from their interactions, as well as the explicit structural information. To model the multiple associations among users and items, we first link them in a \ac{CDS} graph. To model the structure information of the transferred knowledge, we then develop a domain-aware \ac{GCN} to learn user-specific node representations, where two attention mechanisms are devised to weight the local neighbors of the target.
To enhance the item and account representation learning, a time interval-aware message passing strategy and an interactive characteristic modeling component are further devised.
The experimental results on two real-world datasets show the superiority of our graph-based solution.

One of the limitations of \ac{TiDA-GCN} is that it limits the number of domains to two, which may not match the real-world application scenarios. As one of our further works, we will extend \ac{TiDA-GCN} to a more practical multi-domain scenario, and deal with the more serious negative transfer issue within it, since the information/knowledge in each domain should be transferred to other domains more than once.

\section*{Acknowledgments}
This work was supported by the National Natural Science Foundation of China (Nos. 61602282, 62072279),  Australian Research Council
(Nos. DP190101985, FT210100624) and China Postdoctoral Science Foundation (No. 2016M602181).

\ifCLASSOPTIONcaptionsoff
  \newpage
\fi



%



\bibliographystyle{IEEEtran}
\bibliography{Reference}

%
\vspace{-1cm}
\begin{IEEEbiography}
[{\includegraphics[width=1in,clip,keepaspectratio]{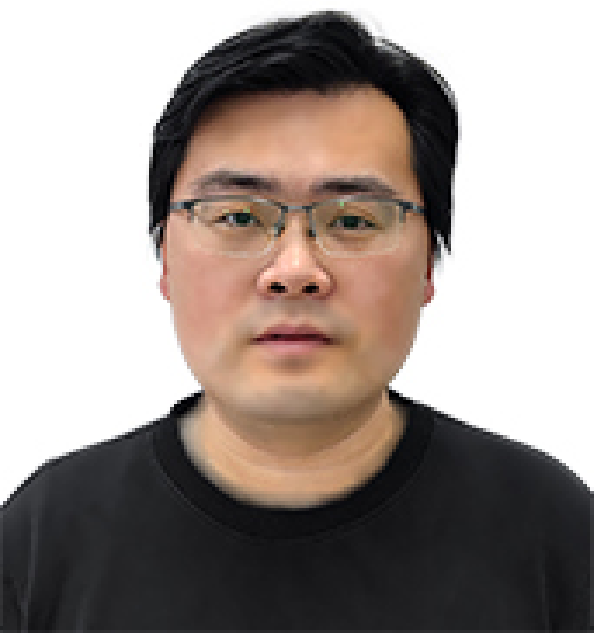}}]
{Lei Guo}
received his Ph.D. degree in computer science from Shandong University, China, in 2015. He is currently an Associate Professor and a Master Supervisor with Shandong Normal University, China. He is a Director of Shandong Artificial Intelligence Society and a member of the Social Media Processing Committee of the Chinese Information Society. His research interests include information retrieval, social networks, and recommender systems.
\end{IEEEbiography}
\vspace{-1.25cm}
\begin{IEEEbiography}
[{\includegraphics[width=1in,clip,keepaspectratio]{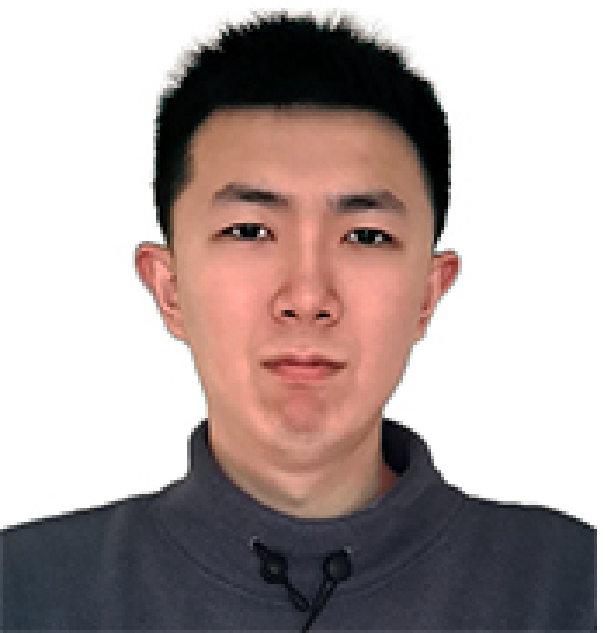}}]
{Jinyu Zhang}
is currently a Master candidate on computer science at the School of information science and Engineering, Shandong Normal University, China. 
His research interests include sequential recommendation and cross-domain recommendation.
\end{IEEEbiography}
\vspace{-1.25cm}
\begin{IEEEbiography}
[{\includegraphics[width=1in,clip,keepaspectratio]{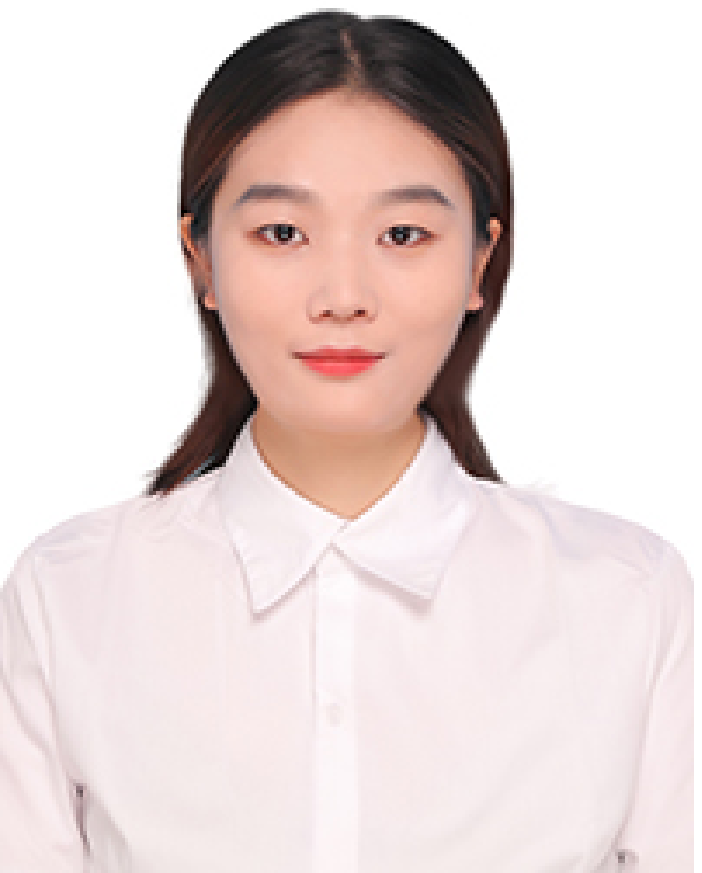}}]
{Li Tang}
is currently a Master candidate on management science and engineering at the School of Business, Shandong Normal University, China. 
Her research interests include cross-domain recommendation and social media mining.
\end{IEEEbiography}
\vspace{-1.25cm}
\begin{IEEEbiography}
[{\includegraphics[width=1in,height=1.25in,clip,keepaspectratio]{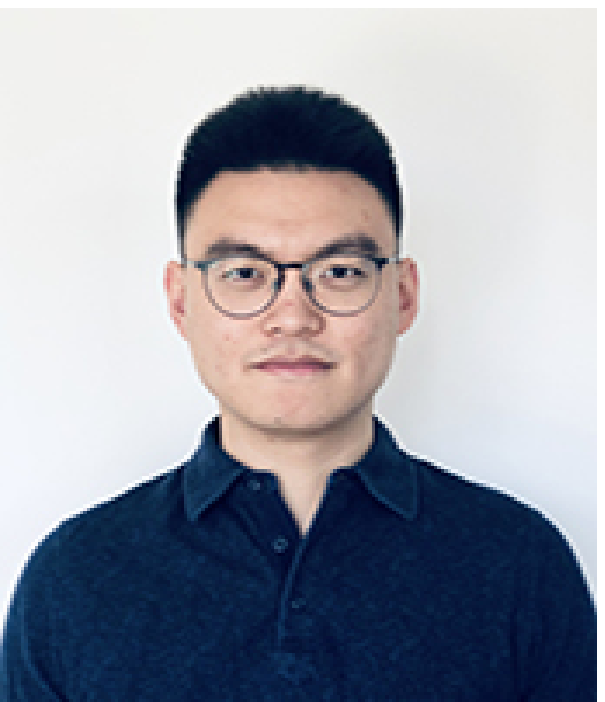}}]
{Tong Chen}
received his Ph.D. degree in computer science from The University of Queensland in 2020.
He is currently a Lecturer with the Data Science research
group, School of Information Technology and Electrical Engineering, The University of Queensland. His research interests include data mining, recommender systems, user behavior modelling and predictive analytics.
\end{IEEEbiography}
\vspace{-1.25cm}
\begin{IEEEbiography}
[{\includegraphics[width=1in,clip,keepaspectratio]{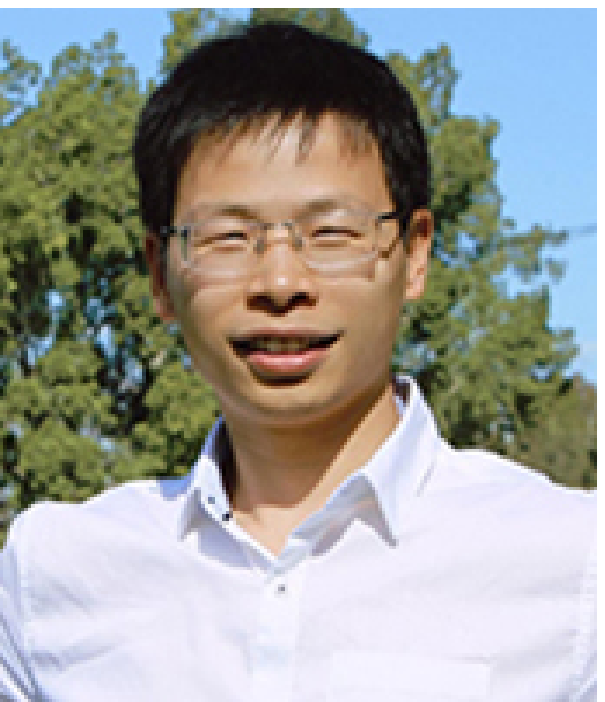}}]
{Lei Zhu}
received his B.S. degree (2009) at Wuhan University of Technology, the Ph.D. degree (2015) at Huazhong University of Science
and Technology. He is currently a full Professor with the School of Information Science and Engineering, Shandong Normal University, China. He was a Research Fellow at the University of Queensland (2016-2017), and at the Singapore
Management University (2015-2016). His
research interests are in the area of large-scale multimedia content analysis and retrieval.
\end{IEEEbiography}
\vspace{-1.25cm}
\begin{IEEEbiography}
[{\includegraphics[width=1in,height=1.25in,clip,keepaspectratio]{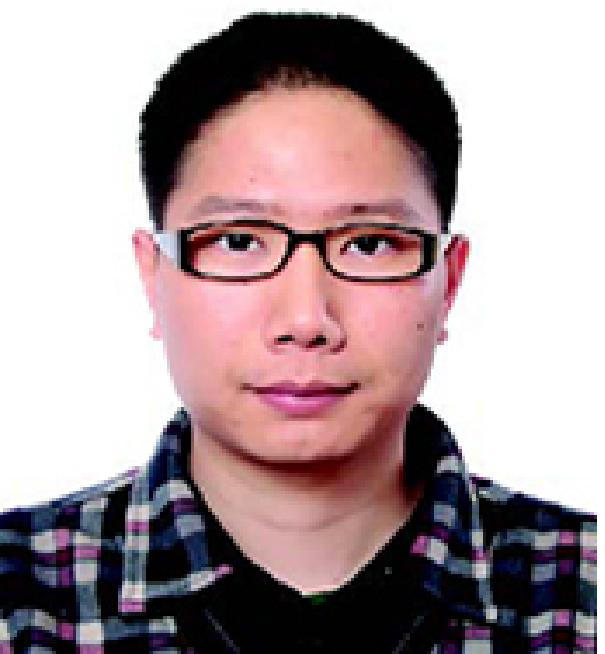}}]
{Hongzhi Yin}
received his Ph.D. degree in computer science from Peking University in 2014. He is an
Associate Professor and Future Fellow with the University
of Queensland. He received the Australian Research Council Future Fellowship and Discovery
Early-Career Researcher Award in 2016 and 2021, respectively. His research interests include recommendation system, user profiling, topic models, deep
learning, social media mining, and location-based services.
\end{IEEEbiography}





\end{document}